\documentclass[%
 reprint,
nofootinbib,
 amsmath,amssymb,
 aps,
 prb,
 superscriptaddress,
]{revtex4-2}
\usepackage{hyperref}
\usepackage{xcolor}
\usepackage{cleveref}
\usepackage{physics}
\usepackage{multirow}
\usepackage{graphicx}
\usepackage{dcolumn}
\usepackage{orcidlink}
\usepackage{bm}
\usepackage{comment}

\begin{document}

\title{
Edge spin generalization for quantum Hall Laughlin droplets with elliptic shapes}

\author{Alexander Fagerlund\orcidlink{0009-0009-3196-0286}}
\email{alexander.fagerlund@fysik.su.se}
 \affiliation{Department of Physics, Stockholm University, AlbaNova University Center, 106 91 Stockholm, Sweden}
 
\author{Alberto Nardin\orcidlink{https://orcid.org/0009-0003-6465-8995}}
\email{alberto.nardin@unitn.it}
\affiliation{Pitaevskii BEC Center, CNR-INO and Dipartimento di Fisica, Universit\`{a} di Trento, Trento, Italy}
 
\author{Leonardo Mazza\orcidlink{https://orcid.org/0000-0002-6754-127X}}
\email{leonardo.mazza@universite-paris-saclay.fr}
\affiliation{Université Paris-Saclay, CNRS, LPTMS, 91405 Orsay, France}
\affiliation{Institut Universitaire de France, 75005 Paris, France}

\author{Eddy Ardonne\orcidlink{https://orcid.org/0000-0003-4479-7106}}
\email{ardonne@fysik.su.se}
 \affiliation{Department of Physics, Stockholm University, AlbaNova University Center, 106 91 Stockholm, Sweden}

\date{\today}

\begin{abstract}
It is known that quasiparticles in the bulk of quantum Hall systems possess an emergent spin, which satisfies a spin-statistics relation. Recent results show that this spin can \textit{fractionalize}: a quasiparticle spin in the bulk induces an opposite edge spin on the boundary. So far, the edge spin has only been studied in the simplest possible geometries, where the boundary is either circular or straight, without considering that a generic confining potential could give the droplet a non-symmetric shape. Yet, the value of the edge spin is not dictated by symmetry, but by topology: it is expected that the specific shape of the edge should not matter. In this work, we therefore generalize the edge spin to elliptically shaped droplets. We consider two different generalizations of the edge spin and test them on a family of elliptic Laughlin states introduced in the present work. The first, most straightforward, generalization is based on the properties of the single-particle orbitals for the elliptic confining potential. We show that this proposal applies to non-interacting states. The second proposal is based on the emergent properties of the quantum fluid. We show that this proposal also applies to interacting states, making it the suitable definition of the edge spin in the elliptic case.
Our work demonstrates that it is possible to define an edge spin for Laughlin fluids with a non-trivial shape. 
\end{abstract}

\maketitle

\section{Introduction}
The fractional quantum Hall (FQH) effect has attracted considerable interest since its original discovery~\cite{tsui1982fqhe}. 
The reasons for this include its link to topological order and topological field theory, as well as the experimental measurements demonstrating that FQH states host anyons with fractional charge and statistics~\cite{girvin1987order,zhang1989eft,halperin1984statistics,arovas1984fractional,wilczek1982spin,moore1991nonabelions,nakamura2020observation,nakamura2022impact,nakamura2023fabry,bartolomei2020fractional,ruelle2023comparing,ruelle2025time}.
Moreover, non-Abelian anyons have been proposed as the basis for error-tolerant qubits \cite{nakamura2020observation,goldman1995tunneling,saminadayar1997charge,kitaev2003computation,nayak2008nonabelian}. 

A series of early works considered the possibility of defining a spin for quasiparticles~\cite{li1992spin,sondhi1992longrange,einarsson1995spin, leinaas1999spinstatisticsquantumhall}, focusing specifically on the anyonic quasiholes in the bulk of FQH systems. 
A later series of articles investigated the link between the quasiparticle conformal dimension and the second moment of the depletion density of the quantum Hall fluid, which is due to the insertion of magnetic flux~\cite{CAN2015752, can2016emergent, schine2019electromagnetic}. Following this, recent work has proposed a definition of a quasiparticle spin~\cite{comparin2022spin} in terms of the depletion density.
This definition is particularly useful because it satisfies a spin-statistics theorem relating the spins of the quasiparticles to their braiding properties~\cite{nardin2023ssr, trung2023ssr}. 
The quasiparticle spins can be computed numerically, which allows conclusions to be drawn about the quasiparticle statistics even when the braiding is non-Abelian. An example of this was presented in Ref.~\cite{fagerlund2025bosonic}, where non-Abelian quasiparticles of the $k=3$ Read-Rezayi state~\cite{read1999parafermions} were studied.

More recently, we showed that the aforementioned spin fractionalizes between bulk and boundary, leading us to propose a definition of a genuine \textit{edge spin} $J_e$ \cite{fagerlund2025spin}. Denoting the spin of the bulk quasiparticle by $J_{qp}$ and setting $\hbar=1$, this fractionalization takes the simple form 
\begin{equation}\label{Eq:Fractionalization}
 J_{qp}=-J_e\;\mathrm{mod}\;1.   
\end{equation}
However, the fractionalization has only been shown for model wavefunctions (including the Laughlin and the Read-Rezayi $k=3$ states)
on a cylinder, where the edge is straight. In these cases, the definition of $J_e$ is related to that of the edge dipole moment in Refs.~\cite{park2014dipole, perkovic2026dipole}. 
When the edge is exactly circular, simple arguments presented in Ref.~\cite{nardin2023ssr} allow an edge spin to be defined based on the notion of a gauge-invariant generator of rotations (GIGR). In this simple case, the GIGR coincides with the angular momentum operator in the symmetric gauge.
In a real experiment, however, the edge profile will never be exactly straight or circular. Therefore, a natural question arises: how does the concept of edge spin generalize to other geometries, where the edge is neither straight nor circular? 

The question of edge spins for other edge geometries also connects to recent work  on deformed edges in quantum Hall systems (see specifically Ref.~\cite{oblak2024anisotropic}), where the influence of the confining potential on the edge has been studied for single-particle states. 
The study of interacting FQHE many-body wavefunctions with deformed edges is non-trivial, and there is some debate on the most appropriate model wavefunctions to employ~\cite{moosavi2025plasma}. 
Our work aims to present a study of the edge spin for both non-interacting and strongly correlated Laughlin-like many-body states forming ellipse-shaped droplets.     
This specific shape was chosen for its simplicity, as it still allows for some semi-analytical results, while being non-trivial and able to unveil remarkable physical effects. 

In this article, we introduce elliptically shaped Laughlin states, which can easily be studied using Monte Carlo techniques based on the plasma analogy. We call these \textit{elliptic Laughlin states}.
In the non-interacting ($\nu = 1$) case, we
show that the external potential confining the droplet into an elliptic shape induces a very natural generalized GIGR. The latter has the elliptic single-particle orbitals in the lowest Landau level (LLL) as eigenfunctions, and is denoted by $J_{\rm orb}$. 
For the non-interacting states composed of these orbitals, $J_{\rm orb}$ provides a proper and natural definition of a fractionalized edge spin.

In order to tackle the interacting case, 
we introduce an additional deformed GIGR, dubbed $J_{\rm int}$, which can be used to define a fractional and quantized edge spin for a generic elliptic Laughlin wavefunction. This can be done regardless of whether the state is interacting or not.

The structure of the paper is as follows. 
In Sec.~\ref{sec:elliptic} we discuss the problem of quantum particles confined in two dimensions, exposed to an orthogonal magnetic field, and subject to elliptic confinement without rotational symmetry. The single-particle orbitals are presented in Sec.~\ref{sec:orbs}, and the many-body elliptic Laughlin states are introduced in Sec.~\ref{sec:many-body}. 
Our focus then shifts to finding the most appropriate GIGR: this is done for single-particle orbitals in Secs.~\ref{lzours}
and~\ref{SubSec:Integral:Kernel}.
In Sec.~\ref{sec:edgenonint}, we show that as expected, this GIGR can be used to define a properly quantized edge spin for the non-interacting elliptic Laughlin states representing a $\nu=1$ integer quantum Hall state. 
On the other hand, the inadequacies of this definition for interacting states are duly reported.

In order to define an appropriate edge spin that also encompasses the interacting case, we establish some algebraic properties between the circular and elliptic Laughlin states in Sec.~\ref{sec:algebraic}. We find a unitary operator deforming the Hamiltonians of the two systems into each other in Sec.~\ref{sec:derivation}, and use this operator to map the circle-shaped single-particle LLL orbitals to the elliptic orbitals in Sec.~\ref{Sec:Def:SP:Orbitals}.
Finally, in Sec.~\ref{SubSec:Deforming:Laughlin}, we show that the elliptic Laughlin states cannot be viewed as a similar deformation of the corresponding circular states.
The algebraic framework used in this latter discussion allows us to find a many-body generator of rotations valid for interacting elliptic states, which is presented in Sec.~\ref{Sec:Laugh:GoodGenerator}.
This generator is based on the geometry of the state but not on any single-particle properties. We utilize this generator of rotations to compute edge spins for a variety of elliptic states, spanning several filling fractions, in Sec.~\ref{sec:edgeint}. 
Our results show that spin fractionalization holds also for elliptically shaped droplets. A critical discussion and conclusions can be found in Sec.~\ref{sec:concl}, together with an outlook.
Some of the more intricate details are relegated to the appendices.

\section{Quantum Hall physics with an elliptic potential}\label{sec:elliptic}

We consider a charged quantum particle in two dimensions in the presence of a perpendicular magnetic field and subject to an elliptic confining potential. This is arguably the simplest situation for studying quantum Hall droplets whose edge shape is more sophisticated than a straight line or a circle. 
From the single-particle LLL orbitals obtained in Sec.~\ref{sec:orbs}, which have a deformed elliptic shape, several elliptic many-body states can be defined. We do this in Sec.~\ref{sec:many-body}. Our states generalize the model wavefunctions originally presented by Laughlin~\cite{laughlin1983qhe}. 

\subsection{LLL Orbitals}\label{sec:orbs}
We study an electron subject to the single-particle Hamiltonian $H=H_K+V_{\rm ell}$.
Here, $H_K$ represents the kinetic term in the presence of an external magnetic field, which is included via the mechanical momenta $\pi_x$ and $\pi_y$ obeying $[\pi_x,\pi_y]=i\frac{\hbar^2}{\ell_B^2}$, and
$V_{\rm ell}$ is the confining potential:
\begin{equation}\label{pot}
    H_K=\frac{\pi_x^2+\pi_y^2}{2m_e}   ,
    \qquad 
    V_{\rm ell}(x,y)=\frac{\mu}{2}\big(\omega_x^2x^2+\omega_y^2y^2\big);
\end{equation}
we assume $\mu>0$ and $\omega_x, \omega_y>0$. The momenta are given by
\begin{equation}
    \mathbf{\pi}_x = -i\hbar \frac{\partial}{\partial x} - A_x,\quad \pi_y = -i\hbar \frac{\partial}{\partial y} - A_y,    
\end{equation}
and we use the symmetric gauge $\mathbf{A}=\frac{B}{2}(-y,x,0)^T$. 
Note that $[\pi_x,x]=[\pi_y,y]=-i\hbar$, and hence the two terms in the Hamiltonian do not commute.

We find the LLL orbitals by introducing the ladder operator $a=\frac{\ell_B}{\sqrt{2}\hbar}(\pi_x+ i\pi_y)$ and its Hermitian conjugate $a^\dagger$, which
satisfy $[a, a^\dagger]=1$.
This can be used to rewrite the kinetic term in the Hamiltonian as 
\begin{equation}\label{HK}
    H_K=\frac{\hbar^2}{m_e\ell_B^2}\bigg(a^{\dagger}a+\frac{1}{2}\bigg).    
\end{equation}
The LLL orbitals are then annihilated by $a$. This leads to the well-known solution $\psi(z,\bar{z})=f(z)e^{-\frac{|z|^2}{4\ell_B^2}}$ for a holomorphic function $f(z)$ with $z=x+iy$ and $\bar{z}=x-iy$.
By projecting any confining potential $V(x,y)$ to the LLL and diagonalizing the potential energy one ends up making a specific choice of orbitals, and hence determining the functions $f(z)$ uniquely.

In terms of the guiding center coordinate operators
\begin{equation}\label{centers}
    R_x=x+\frac{\ell_B^2}{\hbar}\pi_y,\quad R_y=y-\frac{\ell_B^2}{\hbar}\pi_x,
\end{equation}
the LLL projected potential becomes (projection is denoted by a tilde)
\begin{align}\label{vlllcoord}
    \tilde {V}_{\rm ell}(x,y)=\frac{\mu}{2} \left[\bigg(\omega_x^2R_x^2+\omega_y^2R_y^2\bigg)
    +\frac{\ell_B^2}{2}\bigg(\omega_x^2+\omega_y^2\bigg) \right].
\end{align}
This projection procedure is valid when the confinement energy is small compared to the energy gap between Landau levels. Geometrically, this corresponds to a situation where the magnetic length $\ell_B$ is smaller than the typical length scale over which the potential varies.

\begin{figure}[t]
    \centering
        \includegraphics[trim={1cm 1.5cm 0 2.5cm},clip,width=0.9\columnwidth]{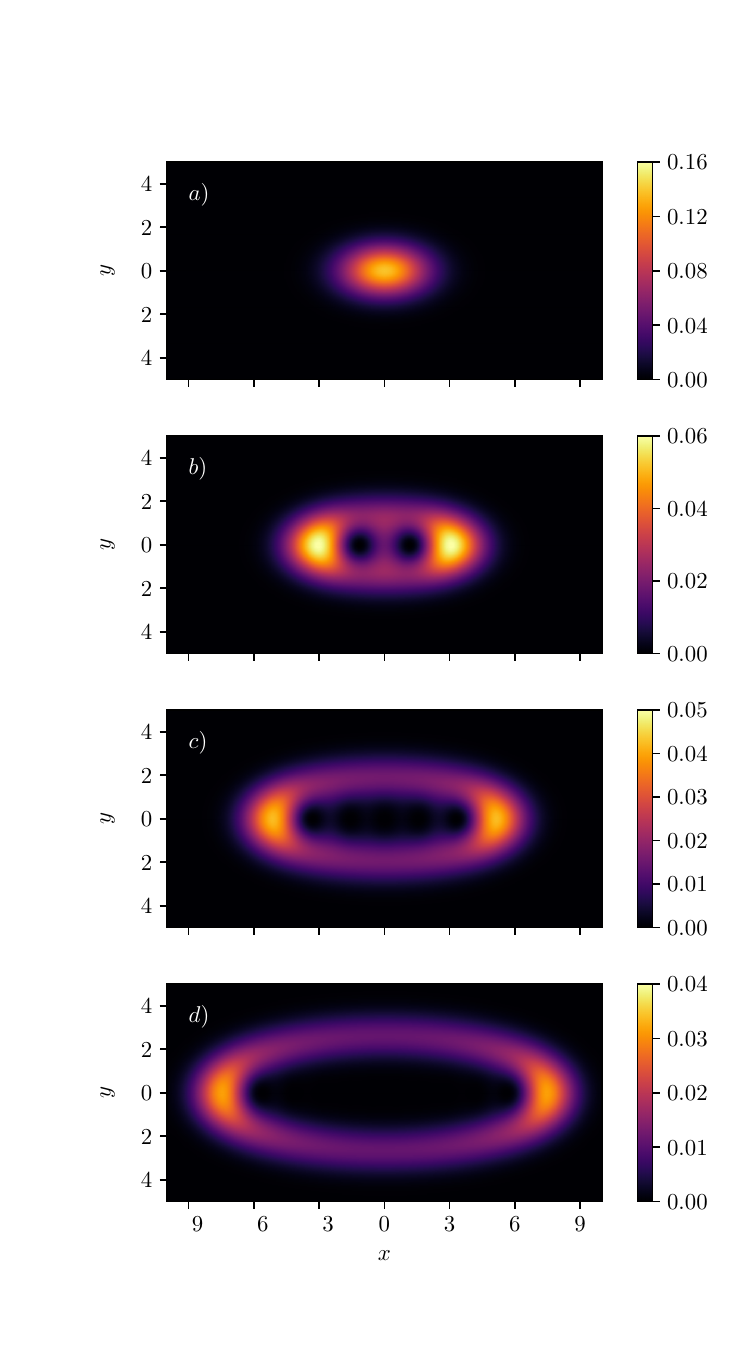}
    \caption{The probability densities $|\psi_{m}(x,y)|^2$ for $\omega_x=1,\omega_y=3$ and a) $m=0$, b) $m=2$, c) $m=5$ and d) $m=10$, with $\psi_{m}$ given by \eqref{orbs}.
    }
    \label{fig:orbex}
\end{figure}

Solving the eigenvalue problem (see App. \ref{App:Landau} for details) gives the eigenstates
\begin{equation}\label{orbs}
    \psi_m=C_mH_m\left(\zeta \right)
    e^{\epsilon\frac{z^2}{4\ell_B^2}}
    e^{-\frac{|z|^2}{4\ell_B^2}}, \quad \text{with} \;
    \zeta = \frac{z}{\ell_B} \sqrt{\frac{\omega_x \omega_y}{\omega_y^2 - \omega_x^2}}.
\end{equation}
Here $H_m$ is the $m$th Hermite polynomial, and the coefficient
\begin{equation}\label{coeffs}
    C_m=\sqrt{\frac{\sqrt{1-\epsilon^2}}{2\pi \ell_B^2}}\frac{\epsilon^{\frac{m}{2}}}{\sqrt{2^mm!}},
    \quad \text{with} \;\; 
    \epsilon=\frac{\omega_y-\omega_x}{\omega_x+\omega_y},
\end{equation}
gives a normalized state. 
The parameter $\epsilon$ equals the so-called \textit{third flattening} of the ellipse-shaped level curves of the potential, see Ref.~\cite{karney2024auxiliary}.
The ellipticity $e$ is related to $\epsilon$ via $e^2=\frac{4\epsilon}{(1+\epsilon)^2}$ and
$\epsilon=0$ if and only if the potential is circularly symmetric. In that case, the orbitals~\eqref{orbs} reduce to the usual LLL orbitals $\phi_m \propto z^me^{-\frac{|z|^2}{4\ell_B^2}}$. 

Figure~\ref{fig:orbex} shows a set of probability densities associated with the orbitals~\eqref{orbs}.
It can be seen that they are shaped like ellipses for $\epsilon\neq 0$, i.e.~for a non-circular potential where $\omega_x\neq \omega_y.$ We also note the angle-dependent “rollercoaster" behaviour of the density, which was extensively discussed in the large-$m$ limit in Ref.~\cite{oblak2024anisotropic} but is seen here to occur also for the exact small-$m$ orbitals. 
\newline
 \subsubsection{Relation to previous work}
  
A version of the calculation of the elliptical orbitals appeared in Ref.~\cite{qiu2012anisotropic}, with the anisotropy entering through an effective mass tensor. 
We point out that the anisotropic Gaussian in the density $|\psi_m|^2\propto e^{-\frac{1}{2\ell_B^2}\big((1-\epsilon)x^2+(1+\epsilon)y^2\big)}$ is the same as in their result, which becomes apparent after an appropriate change of coordinates. 
Another closely related calculation was performed in Ref.~\cite{oblak2024anisotropic}, where the authors find approximate eigenstates of LLL-projected anisotropic potentials. The solutions are labeled by a large quantum number $m$, which the authors view as a generalized angular momentum. We interpret $m$ similarly in App.~\ref{sec:lzprx}. 

Additionally, our study of anisotropic wavefunctions connects to the work of Ref.~\cite{mizrahi1992squeezed}, where squeezed harmonic oscillator wavefunctions are found using generalized Hermite polynomials. Anisotropic harmonic oscillators have also been studied in Refs.~\cite{patra20232dho,rebane1972anisotropic}. 
In Ref.~\cite{muhuri2021entanglement}, the anisotropic harmonic oscillator is solved by similar algebraic techniques as 
in Sec.~\ref{sec:orbs} (and App.~\ref{App:Landau}). The authors find a squeezed Gaussian ground state similar to our orbital $\psi_0$ in Eq.~\eqref{gs}. 

For additional related research, we refer to Refs. \cite{madhav1994electronic,Qiong-Gui2002anisotropic,aftalion2009fast,ghazanfari2010rapidly}.

\subsection{Elliptic Laughlin states }\label{sec:many-body}

Having obtained the single-particle LLL orbitals, we now construct various many-body states of the system. To begin with, a many-fermion state representing a non-interacting integer quantum Hall system can be found by considering a Slater determinant of the orbitals~\eqref{orbs}:
\begin{widetext}
\begin{align}\label{iqhfun}
    \Psi_{\nu=1} (z_i) &=\begin{vmatrix}
H_0(\zeta_1)  & H_0(\zeta_2)     & ...      &   H_0(\zeta_{N})       \\
H_1(\zeta_1)  & H_1(\zeta_2)     & ...      &   H_1(\zeta_{N})  \\
\vdots & \vdots & \ddots & \vdots    \\
H_{N-1}(\zeta_1)  & H_{N-1}(\zeta_2)     & ...      &   H_{N-1}(\zeta_{N})     
\end{vmatrix}e^{-\frac{1}{4\ell_B^2}\sum_k|z_k|^2}e^{\frac{\epsilon}{4\ell_B^2}\sum_lz_l^2}\nonumber\\
&\propto\begin{vmatrix}
1  & 1     & ...      &   1       \\
z_1  & z_2     & ...      &   z_{N}  \\
\vdots & \vdots & \ddots & \vdots    \\
z_1^{N-1}  & z_2^{N-1}     & ...      &   z_{N}^{N-1}     
\end{vmatrix}e^{-\frac{1}{4\ell_B^2}\sum_k|z_k|^2}e^{\frac{\epsilon}{4\ell_B^2}\sum_lz_l^2}\propto\prod_{i<j}(z_i-z_j)e^{-\frac{1}{4\ell_B^2}\sum_k|z_k|^2}e^{\frac{\epsilon}{4\ell_B^2}\sum_lz_l^2}.
\end{align}
\end{widetext}
The second step follows from elementary row operations for determinants and the structure of the Hermite polynomials, and is an equality up to an unimportant overall factor. This step is valid as long as orbital $j$ has a leading term of degree $j$ for all the values $j=0,1,...,N-1$, and thus holds more generally than just for the Hermite polynomials.
In the rotationally symmetric case, characterized by $\epsilon=0$, the state \eqref{iqhfun} reduces to the well-known Slater determinant of the circular orbitals typically discussed in lecture notes~\cite{tong2016lecturesquantumhalleffect,goerbig2009quantumhalleffects}. 

\begin{figure}
\centering
        \includegraphics[trim = {2.5cm 0.5cm 0.5 2.5cm}, clip, width=1.\columnwidth]{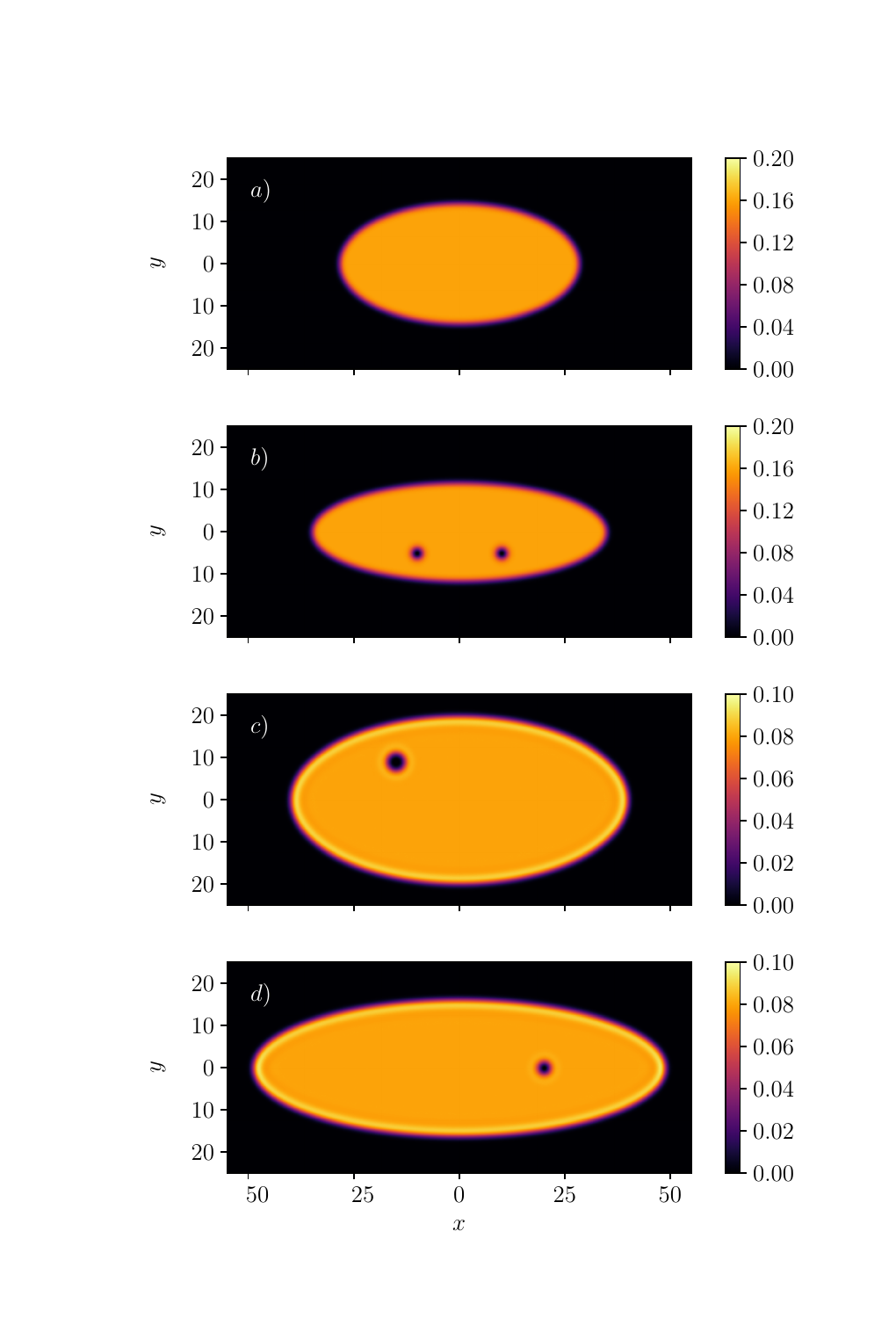}
    \vspace{-1 cm}
     \centering
    \caption{Metropolis sampled densities for systems with $N=200$ particles. In all panels, $\omega_x=1.$ The droplets pictured have a) no quasiholes, $M=1,\omega_y=2$, b) two quasiholes, $M=1,\omega_y=3$, c) a double quasihole (i.e. two quasiholes at the same position), $M=2,\omega_y=2$ and d) one quasihole, $M=2,\omega_y=3$. The bulk density depends only on $M$ and equals $\frac{\nu}{2\pi}=\frac{1}{2\pi M}$. Since $N$ is fixed, the area of the droplet scales with $M$. 
    }
    \label{fig:shifted}
\end{figure}

It is natural to generalize our wavefunction \eqref{iqhfun} to include a quasihole at $w$ as follows:
\begin{equation}\label{Eq:IQH:qh}
    {\Psi}_{\nu=1, {\rm qh}} (z_i; w) = \prod_k(z_k-w) \times \Psi_{\nu=1}(z_i).
\end{equation}
One may also generalize $\Psi_{\nu=1}(z_i)$ to an interacting system with filling fraction $\nu=1/M$, for integer $M>1$. This gives an ellipse-shaped analogue of the $\nu=1/M$ Laughlin state:
\begin{equation}\label{lllfun}
    \Psi_{\nu=\frac{1}{M}}(z_i)=\prod_{i<j}(z_i-z_j)^Me^{-\frac{1}{4\ell_B^2}\sum_r|z_r|^2}e^{\frac{\epsilon}{4\ell_B^2}\sum_sz_s^2}.
\end{equation}
Finally, for completeness we also point out that when the filling fraction is $\nu=1/M$ and the system hosts a quasihole at $w$, the wavefunction we consider is
\begin{equation}\label{qhfun}
   { \Psi}_{\nu=\frac{1}{M}, {\rm qh}} (z_i; w)=\prod_k(z_k-w) \times \Psi_{\nu=\frac{1}{M}}(z_i).
\end{equation}
The states in Eqs.~\eqref{lllfun} and~\eqref{qhfun} can be derived by multiplying the corresponding circular Laughlin state with the anisotropic exponential factor $e^{\frac{\epsilon}{4 \ell_B^2} \sum_s z_s^2}$.
We elaborate on this from an operator perspective in Sec.~\ref{SubSec:Deforming:Laughlin}.

We have seen that the elliptic states at $\nu=1$ are the many-body ground states of the Hamiltonian $H_K+ V_{\rm ell}$, in the limit of large magnetic field and weak confinement. However, it is currently unknown whether the wavefunction~\eqref{lllfun} for $M>1$ is the ground state of some appropriate interacting Hamiltonian in the presence of our elliptic confinement $V_{\rm ell}$. 

When rotational symmetry is broken, i.e. $\epsilon\neq 0$, the droplets described by Eqs.~\eqref{iqhfun}-\eqref{qhfun} become elliptic in shape. This can be seen through a standard sampling of the density using a Monte Carlo algorithm. 
Indeed, it is well-known that the density of the Laughlin wavefunction can be rewritten as the partition function $e^{-\beta E}$ of a two-dimensional plasma \cite{laughlin1983qhe}. One easily sees that the density of our state \eqref{lllfun} can be written in the same way, except with the replacement 
\begin{equation}\label{plasmaexp}
    \sum_{i=1}^N\frac{x_i^2+y_i^2}{2\ell_B^2}\rightarrow \sum_{i=1}^N\frac{(1-\epsilon)x_i^2+(1+\epsilon)y_i^2}{2\ell_B^2}
\end{equation}
in the energy in the exponent of the partition function. Thus, our state is also described by a plasma, but with ellipse-shaped confinement.

Some results are presented in Fig.~\ref{fig:shifted}, which contains example densities both with and without quasiholes. These densities are computed using Metropolis sampling \cite{metropolis1953metropolis}.
The case $\epsilon>0$ corresponds to a density stretched in the $x$ direction, which is to be expected since the confinement is stronger along $y$. This can be seen in Fig. \ref{fig:shifted}.
Because $-1<\epsilon <1$ for $0<\omega_x,\omega_y<\infty$, the Gaussian factor still gives the proper decay at infinity. 
Despite the droplet being stretched, the bulk is still perfectly homogeneous. On the other hand, the shape and properties of the \textit{edge} change. This is seen in Fig. \ref{fig:shifted}, for different $\epsilon$ and quasihole configurations.

For later convenience, the states in Eqs.~\eqref{iqhfun}-\eqref{qhfun} are referred to as \textit{elliptic Laughlin states.}
In Appendix~\ref{App:BulkProperties} and~\ref{app:bulk}, the bulk properties of these states are extensively discussed with the help of numerical simulations. We find that the bulk of the elliptic Laughlin state is completely indistinguishable from that of the widely studied circular droplets. For instance, the quasihole density profiles are isotropic.
On the other hand, the edge differs. Based on the different edge structure, we conjecture that our elliptic Laughlin state for $\nu<1$ is a low-energy excited state of the interacting QH Hamiltonian projected to the LLL, with contact interactions and elliptic confinement.

\subsubsection{Relation to previous work}

To the best of our knowledge, this is the first work to study the wavefunctions in Eqs.~\eqref{iqhfun}-~\eqref{qhfun}. 
Anisotropic Laughlin states were investigated analytically and numerically before~\cite{yang2012band,krishna2019interaction}, and ellipse-shaped versions closely resembling our state~\eqref{lllfun} exist in the literature~\cite{qiu2012anisotropic,johri2016probing,wang2012anisotropic}. Laughlin fluids (and modifications thereof) with deformed correlation functions have also been considered \cite{haldane2011geometrical}, including systems with broken rotational symmetry \cite{ musaelian1996broken, ciftja2001hypernetted, ciftja2004liquid}. 
Nevertheless, previous works on ellipse-shaped states did not discuss the density profiles and bulk spins of quasiholes in these states, although we note that pair correlation functions were computed. 
Moreover, none of these papers study the edge spin of their anisotropic states, which is the goal of the present work. We note, however, that Ref.~\cite{haldane2011geometrical} contains a quantized “guiding center spin" associated with correlation functions of the system. 

\begin{figure}
    \centering
    \includegraphics[width=\linewidth]{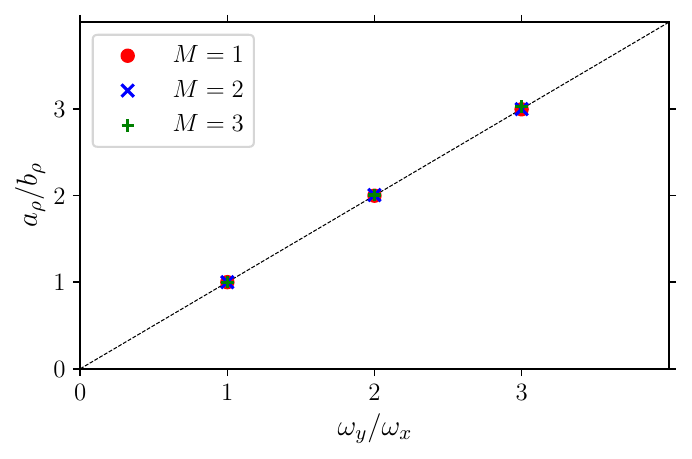}
    \caption{Ratio $a_{\rho}/b_{\rho}$ of the semi-axes $a_{\rho},b_{\rho}$ of ellipse-shaped droplets, where $a_{\rho}$ ($b_{\rho}$) is the semi-major (semi-minor) axis. The ratio is given for various ratios $\omega_y/\omega_x$ of the confinement parameters, and for several different $M$ in the filling fraction $\nu=1/M$. The density was obtained using Monte Carlo sampling. To calculate the axis lengths, we chose the reference value of the density to be half the bulk density, $\frac{1}{2}\frac{\nu}{2\pi}=\frac{1}{4\pi M}$. The dashed black line is the line $a_{\rho}/b_{\rho}=\omega_y/\omega_x.$}
    \label{fig:axes}
\end{figure}

The plasma subject to ellipse-shaped confinement has been studied before~\cite{dean2025two,forrester1996two,di1994laughlin}. 
In these works, the authors find that the axis lengths $a_{\rho},b_{\rho}$ of the plasma obey 
$\frac{a_{\rho}}{b_{\rho}}=\frac{\omega_y}{\omega_x}$, where we have expressed their results in terms of our parameters. As can be verified using Metropolis sampling (see Fig.~\ref{fig:axes}), our droplet also has $\frac{a_{\rho}}{b_{\rho}}=\frac{\omega_y}{\omega_x}$. Therefore, our densities seem consistent with these previous results.

\section{Gauge-invariant generator of rotations for single orbitals}\label{sec:singleorb}

Given the elliptic orbitals $\psi_m$ in Eq.~\eqref{orbs}, three questions are natural to consider. The first is to find a generalized GIGR appropriate for this elliptic geometry, and to examine how it acts on the single-particle orbitals $\psi_m$. The eigenvalues of the GIGR can then be interpreted as generalized angular momenta, which are expected to reduce to the usual ones in the circular limit. The second question is to investigate how the LLL projection of the GIGR can be represented using an integral kernel $K(x,y)$, so that the integral
\begin{equation}\label{jgeneric}
    J=\int K(x,y)|\psi_m(x,y)|^2 dxdy
\end{equation}
acts like the GIGR on the LLL orbitals $\psi_m$. This is a useful question to ask, since such a kernel can be used for the computation of generalized angular momenta by only sampling the density of the system. 
Similar integral kernels have allowed the computation of the spins of anyons satisfying a spin-statistics theorem in the bulk of many-body states~\cite{comparin2022spin,nardin2023ssr,fagerlund2025bosonic}.
Finally, the third question is whether the generator of rotation for single-particle orbitals is guaranteed to be valid also for a many-body system. 

The first two questions above will be answered for ellipse-shaped droplets in Sec.~\ref{lzours} and Sec.~\ref{SubSec:Integral:Kernel}, respectively.
The third question is addressed in Sec.~\ref{sec:edgenonint},
where we see that the kernel derived in Sec.~\ref{lzours} is valid only for non-interacting elliptic many-body systems.

\subsection{GIGR for circular and ellipse-shaped confinement}\label{lzours}

We now proceed to discuss the GIGR for the orbitals~\eqref{orbs}.
We first outline the circularly symmetric case, where $\omega_x=\omega_y$, and then generalize our treatment to find a GIGR appropriate for elliptic LLL orbitals.  

By recalling the commutation relations of the guiding center operators introduced in Eqs.~\eqref{centers}, $[R_x, R_y] = -i\ell_B^2$, and the fact that these operators commute with the mechanical momenta, $[R_i, \pi_j]=0$ for $i,j = x,y$,
it follows that the object
\begin{equation}\label{jcirc}
    J_{\rm circ}=\frac{\hbar}{2 \ell_B^2}\left(
    R^2_x+R^2_y\right)- 
    \frac{\ell_B^2}{2\hbar}
    \left(\pi_x^2+\pi_y^2
    \right)
\end{equation}
satisfies the commutation relations
\begin{align}\label{comms}
    [J_{\rm circ},\pi_x]&=i\hbar\pi_y,\quad [J_{\rm circ},\pi_y]=-i\hbar\pi_x, \nonumber\\  [J_{\rm circ},R_x]&=i\hbar R_y,\quad [J_{\rm circ},R_y]=-i\hbar R_x.
\end{align}
These are the expected commutation relations for a generator of rotations, and the rotations follow the circular level curves of the confinement, since 
\begin{equation}
    [J_{\rm circ}, x] = i \hbar y, 
    \quad 
    [J_{\rm circ}, y ] = - i \hbar x.
\end{equation}
Because the mechanical momenta and the guiding center operators are gauge-invariant, it follows that $J_{\rm circ}$ is gauge-invariant -- hence the name “GIGR". 

For the ellipse-shaped equipotentials, we introduce a generalized GIGR appropriate for a system in the LLL and confined by an elliptic potential, which reads
\begin{equation}\label{jell}
    J_{\rm orb}=\frac{\hbar}{2 \ell_B^2}\left(\frac{\omega_x}{\omega_y} R^2_x+\frac{\omega_y}{\omega_x} R^2_y\right)-\frac{\ell_B^2}{2\hbar}\left(\pi_x^2+\pi_y^2\right).
\end{equation}
With the definition in Eq.~\eqref{jell}, the commutators generalizing those in Eq.~\eqref{comms} become
\begin{align}\label{commsell}
    [J_{\rm orb},\pi_x]&=i\hbar\pi_y,\;\qquad [J_{\rm orb},\pi_y]=-i\hbar\pi_x, \nonumber\\  [J_{\rm orb},R_x]&=i\hbar\frac{\omega_y}{\omega_x}R_y,\quad [J_{\rm orb},R_y]=-i\hbar\frac{\omega_x}{\omega_y}R_x.
\end{align}

We point out that the part involving $\pi_x,\pi_y$ is the same in both $J_{\rm circ}$ and $J_{\rm orb}$ because $\pi_x$ and $\pi_y$ are related to the Landau orbits and to the kinetic energy term $H_K$ rather than the potential term. The kinetic term can be expressed using the ladder operators $a,a^{\dagger}$ as in Eq.~\eqref{HK}, while (as is seen in App.~\ref{App:Landau}) the LLL-projected potential term is related to another, independent oscillator. This decoupling means that the upper commutation relations in Eqs. \eqref{comms} and \eqref{commsell} are of the same form, and do not depend on whether the level curves of the potential are ellipses or circles.  

In order to find the expressions of $J_{\rm circ}$ and $J_{\rm orb}$ projected onto the LLL, we first observe that the former is a specific case of the latter for $\omega_x = \omega_y$.
Moreover, we note that the combination $\frac{\omega_x}{\omega_y}R_x^2+\frac{\omega_y}{\omega_x} R_y^2$ equals $2\ell_B^2(b^{\dagger}b+\frac{1}{2})$, with the operators $b,b^{\dagger}$ given by 
\begin{align}\label{bbdag}
    b&=\frac{1}{\sqrt{2\omega_x\omega_y\ell_B^2}}(\omega_xR_x-i\omega_yR_y),\nonumber\\
    b^{\dagger}&=\frac{1}{\sqrt{2\omega_x\omega_y\ell_B^2}}(\omega_xR_x+i\omega_yR_y),
\end{align}
and satisfying $[b, b^\dagger] = 1$.
Hence, the first term both in $J_{\rm circ}$ and in $J_{\rm orb}$ can be written as $\hbar\big(b^{\dagger}b+\frac{1}{2}\big)$, where $\omega_x$ and $\omega_y$ are equal in the former case and kept generic in the latter. Similarly, 
the second term equals $-\hbar\big(a^{\dagger}a+\frac{1}{2}\big)$, which can be seen using the identities from Sec. \ref{sec:orbs}. This term then becomes $-\frac{\hbar}{2}$ once it is projected to the LLL. Therefore, the LLL projections of $J_{\rm circ}$ and $J_{\rm orb}$ become
\begin{equation}\label{projJ}
\tilde {J}_{\rm circ} 
=\hbar \, b^{\dagger}b,
\qquad 
\tilde {J}_{\rm orb} 
=\hbar \, b^{\dagger}b.
\end{equation}
Because $\hbar \, b^{\dagger}b$ is the LLL projected generator of rotations, it can be interpreted as an angular momentum operator. It can be shown using the relations in App. \ref{App:Landau} that the orbitals $\psi_m(z)$ in Eq.~\eqref{orbs} are eigenvectors of $\tilde {J}_{\rm orb}$ with eigenvalue $\hbar m$. Since the terms which differ between $J_{\rm orb}\psi_m$ and $\tilde{J}_{\rm orb}\psi_m$ are proportional to $a^{\dagger}a \psi_m=0$, it follows that the orbitals $\psi_m$ are eigenvectors of the GIGR:
\begin{equation}\label{jpsi}
    J_{\rm orb} \psi_m = \hbar m \psi_m.
\end{equation}
Furthermore, since the eigenvalue above coincides with that of the angular momentum operator $\hbar b^{\dagger}b$, we may interpret $\hbar m$ as an angular momentum eigenvalue corresponding to the eigenstate $\psi_m$. 

The relation \eqref{jpsi} is analogous to that of the usual angular momentum for the rotationally symmetric states $\phi_m$, and motivates the interpretation of \eqref{jell} as a version of the GIGR appropriate for the elliptic orbitals in Eq. \eqref{orbs}.
To see this, note that the same procedure is typically followed when working in the symmetric gauge in rotationally invariant setups. There, the generator of rotations acts on the LLL orbitals $\phi_m \propto z^me^{-\frac{|z|^2}{4\ell_B^2}}$ as $L_z=\hbar(z\partial_z-\bar{z}\partial_{\bar{z}})$ with eigenvalue $\hbar m$. The generator of rotations is interpreted as an angular momentum operator, and $\hbar m$ as the corresponding angular momentum eigenvalue. Thus, it is natural to interpret $J_{\rm orb}$ and $\hbar m$ analogously.

\subsection{The integral kernel}
\label{SubSec:Integral:Kernel}
The problem of finding the integral kernel stated in Eq.~\eqref{jgeneric} can be reformulated as follows: is there an operator $K_{\rm orb}(x,y)$ depending only on the position operators $x$ and $y$ but not on the canonical momenta $p_x$ and $p_y$
such that, for an LLL state $\ket{\psi}$, the relation $\bra{\psi} K_{\rm orb} \ket{\psi} = \bra{\psi} J_{\rm orb} \ket{\psi}$ is true?

To answer the question, we observe that from the previous relations, the LLL-projected generator of rotations $\tilde{J}_{\rm orb}$ obeys  
\begin{equation}
    \tilde{J}_{\rm orb}=\hbar b^{\dagger}b=\frac{\hbar}{2\omega_x\omega_y \ell_B^2}\bigg(\omega_x^2R_x^2+\omega_y^2R_y^2-\omega_x\omega_y \ell_B^2\bigg). 
\end{equation}
By using Eq.~\eqref{vlllcoord}, we have that within the LLL, $\omega_x^2R_x^2+\omega_y^2R_y^2=\omega_x^2 \tilde x^2+\omega_y^2 \tilde y^2-\frac{\ell_B^2}{2}(\omega_x^2+\omega_y^2)$, where we remind the reader that $\tilde x$ and $\tilde y$ are the position operators projected onto the LLL. 
Therefore, within the LLL, we may write
\begin{equation}\label{repLLL}
    \tilde{J}_{\rm orb}
    =\hbar\bigg(\frac{\omega_x^2 \tilde x^2+\omega_y^2 \tilde y^2}{2\omega_x\omega_y \ell_B^2}-\frac{(\omega_x+\omega_y)^2}{4\omega_x\omega_y}\bigg).
\end{equation}
For this reason, it is natural to propose the following integral kernel: 
\begin{equation}\label{ourkernel}
    K_{\rm orb}(x,y)=\hbar \left(\frac{\omega_x^2x^2+\omega_y^2y^2}{2\omega_x\omega_y\ell_B^2}-
    \frac{(\omega_x+\omega_y)^2}{4\omega_x\omega_y} \right) .
\end{equation}

It is easily verified with an explicit calculation that
$
   \int K_{\rm orb} (x,y)|\psi_m|^2 dxdy=\hbar m
$.
The eigenvalues $\hbar m$ of the generator of rotations $J_{\rm orb}$ are thus reproduced and the operator, once projected onto the LLL, can be represented as a functional of the probability density of the orbital. 
This idea will appear repeatedly in the present work, and generalizes 
the kernel used in Refs.~\cite{comparin2022spin,nardin2023ssr,fagerlund2025bosonic} for rotationally symmetric droplets,
\begin{align}\label{bulkkernel}
    K_{\rm circ}(r,\phi)=
    \hbar \left(
    \frac{r^2}{2\ell_B^2}-1 \right).
\end{align}

Let us now interpret our result. From the previous discussion, the kernel $K_{\rm orb}(x,y)$ coincides with the LLL projection of our generalized generator of rotations for the elliptic confinement in Eq.~\eqref{pot}. Remarkably, the first term in Eq.~\eqref{ourkernel} exactly follows the shape of the potential $V(x,y)$. The second term can be rewritten as 
\begin{equation}
    \frac{(\omega_x+\omega_y)^2}{4\omega_x\omega_y}=\left(\frac{\frac{\omega_x+\omega_y}{2}}{\sqrt{\omega_x\omega_y}}\right)^2.
\end{equation}
The numerator is just the arithmetic mean of the potential parameters $\omega_x,\omega_y$, and the denominator is their geometric mean. Because the arithmetic mean of two positive numbers is always greater than or equal to their geometric mean, with equality if and only if the numbers are equal, the constant term is a measure of the ellipticity of the potential: the further away from $1$ the term is, the more elliptic the equipotential curves are.

We also note that the kernel $K_{\rm orb}$ can be interpreted as an action coordinate in the sense of Ref.~\cite{moosavi2025plasma}. This suggests that Eq.~\eqref{ourkernel} can be generalized to other potential geometries by finding the action coordinates for the confinement in question.
\subsection{Edge spin using the single-particle generator of rotations}\label{sec:edgenonint}

As shown in our previous work, the presence of a quasihole in the bulk of the QH fluid leads to spin fractionalization \cite{fagerlund2025spin}. More precisely, the bulk spin of the quasihole $J_{qp}$ induces an edge spin $J_e$ such that $J_{qp}=-J_e\;\mathrm{mod}\;1$, in units where $\hbar=1$. This was anticipated in Eq.~\eqref{Eq:Fractionalization}, and parallels the well-known fractionalization of charge between a quasihole with charge $q$ and the edge, to which a charge $-q$ is pushed. 
However, this spin fractionalization has been tested only in cases where the edge is either exactly circular or straight~\cite{nardin2023ssr,fagerlund2025spin}. Examining it for elliptic edges is the goal of the present work.

We wish to compute $J_{qp}$ and $J_e$ for the elliptic Laughlin states introduced in Sec.~\ref{sec:many-body}, comparing the situation with a quasiparticle against that without, and investigate whether spin fractionalization holds also when the droplet is deformed. 
Since the quasihole density profiles are the same as for rotationally symmetric confinement, it is expected that the bulk quasihole spins are unchanged. These matters are investigated in App.~\ref{App:BulkProperties}, where we show that the quasiparticle has a circular shape, and in App.~\ref{app:bulk}, where we demonstrate that $J_{qp}$ can be computed using the rotationally symmetric kernel in Eq.~\eqref{bulkkernel}~\cite{comparin2022spin}.

To compute the many-body edge spin, 
we introduce the depletion density $\delta \rho = \rho_{qh} - \rho_0$, namely the difference between the density of the droplet with the quasiparticle $\rho_{qh}$, associated with the wavefunction $\Psi_{\nu = \frac 1M, qh}$ in Eq.~\eqref{qhfun}, and the density $\rho_0$ associated with the wavefunction $\Psi_{\nu = \frac 1M}$ in Eq.~\eqref{lllfun}. We remark that $\delta \rho$ is significantly different from zero only around the quasiparticle and at the edge.
It is natural to define the edge spin for the elliptic droplet $J_e^{\rm orb}$ using the kernel $K_{\rm orb}(x,y)$, which was shown to work for elliptic orbitals:
\begin{widetext}
\begin{align}\label{Jenonint}
    J_e^{\rm orb}(R_{<}) = \hbar
    \int_0^{2 \pi} \hspace{-0.2cm}
    \int_{r=R_<}^{r=R_{>}}\bigg(\frac{r^2\big(\omega_x^2\cos^2(\phi)+\omega_y^2\sin^2(\phi)\big)}{2\omega_x\omega_y\ell_B^2}-\frac{(\omega_x+\omega_y)^2}{4\omega_x\omega_y}\bigg)\delta\rho\;rdrd\phi.
\end{align}
\end{widetext}
The integration is from an inner radial cutoff $R_<$ to a long-distance radial cutoff $R_{>}$ chosen to be well outside the droplet, where the density vanishes, so that the entire edge structure is considered. From now on, $R_{>}$ will be taken at infinity. 
The value $R_<$ should be large enough to avoid including the quasiparticle, but small enough to consider the entire edge.

\begin{figure*}[t]
    \centering
    \includegraphics[scale=1.2, trim = {0.2cm 0cm 0cm 0cm}, clip]{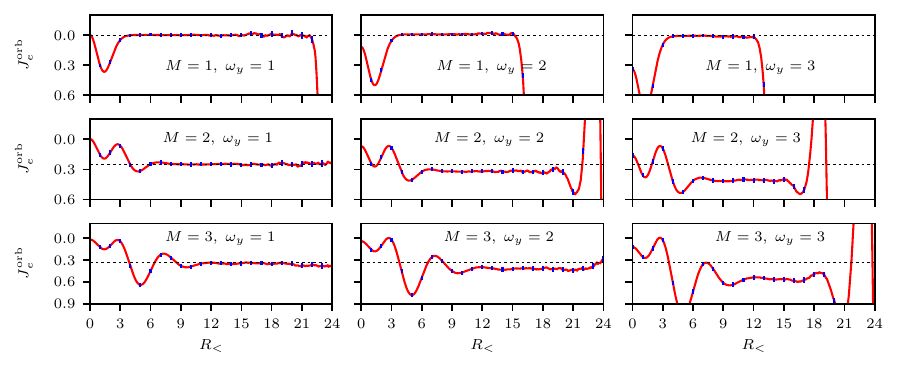}
    \caption{%
    Values of $J_e^{\rm orb}$ for quantum Hall droplets with $\nu=1/M$ for $M=1$ (where $J_e^{\rm orb}$ is applicable) and $M=2,3$ (where it is not, unless the droplet is circular). $J_e^{\rm orb}$ has been computed according to \eqref{Jenonint}, and we have set $\hbar=1$. For each filling fraction, the confinement has $\omega_x=1$ and $\omega_y=1,2$ or $3$. The number of particles is $N=300$. Throughout, the integer $N=300$ has been subtracted, and lengths are in units of $\ell_B$. Edge spins $J_e^{\rm orb}$ are plotted as functions of the inner cutoff $R_{<}$. In the cases where $J_e^{\rm orb}$ is applicable (and only in these cases), the expected spin values $J_e=-\frac{1}{2}(1-\nu)$ (dashed lines) are found in the plateau regions, away from the quasihole at the origin (leftmost region) and the closest parts of the edge (rightmost region). Blue error bars have been included for every $20$th plot point, i.e., for the points where $R_{<}$ takes integer values. Note that the scale differs between different rows but stays the same within each row.}
    \label{fig:jedgenonint}
\end{figure*}

\begin{figure}
    \centering
    \includegraphics[width=\linewidth]{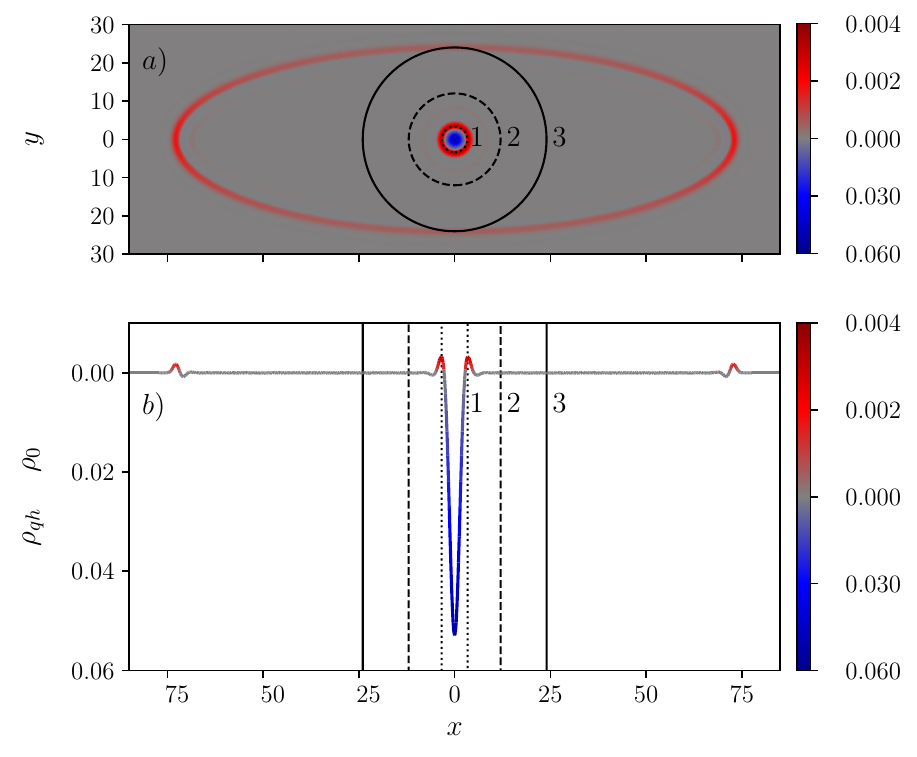}
    \caption{Illustration of the different zones of the FQHE droplet, using Metropolis data for $\delta \rho = \rho_{qh}-\rho_0$ for a system with $N=300$ particles, $\omega_x=1,\omega_y=3$ and $M=3.$ a) The depletion density $\delta \rho(x,y)$. The circles denoted $1$ and $2$ are in the quasihole region and in the unperturbed bulk, respectively. In the upper and lower parts of the figure, circle $3$ is seen to reach the edge region. Note that the color map exaggerates positive values to make the plot easier to read.
    b) A cross-section of the same depletion density $\delta \rho(x, y=0)$, along the $x$ axis. The numbers and the vertical lines denote the same regions as in panel (a).}
    \label{fig:zones}
\end{figure}

Our results are presented in Fig.~\ref{fig:jedgenonint}.
We study the filling fractions $\nu=1/M$ for $M=1,2,3$, and for each filling we consider $\omega_x=1$ and $\omega_y=1,2,3$. 
Details on the computations, and information on how the error bars were produced, are given in Appendix~\ref{app:metropolis}.
In all nine panels, the graph oscillates for small values of $R_<$ because the integration limit $R_<$ moves over the density profile of the quasihole.
For illustration purposes, this region is denoted by $1$ in the sketch Fig.~\ref{fig:zones}.
After the oscillations, the curves in Fig.~\ref{fig:jedgenonint} stabilize: $R_<$ is entirely in the bulk region, where $\delta\rho=0$ up to exponentially suppressed corrections; see region $2$ in Fig.~\ref{fig:zones}. 
As expected, the oscillations vanish more rapidly for the non-interacting ($M=1$) states than for the interacting ones ($M>1$), 
because the correlation length of the fluids increases with $M$.

These plateaux are taken to be the edge spin values $J_e^{\rm orb}$ of the respective fluids, \textit{as predicted by the single-particle kernel.}
Finally, $R_<$ is further increased and reaches the edge region, corresponding to region $3$ in Fig.~\ref{fig:zones}. This causes high-amplitude oscillations as $\delta\rho$ deviates from $0$. 
The transitions from oscillations to a plateau and then back to oscillations again is generic and expected: see Refs.~\cite{comparin2022spin,nardin2023ssr,fagerlund2025bosonic,fagerlund2025spin}.

In addition, there are some trends worth pointing out in Fig.~\ref{fig:jedgenonint}. Keeping $M$ fixed and increasing $\omega_y$ makes the plateau shorter. The reason is that increasing $\omega_y$ with $\omega_x=1$ held constant causes the droplet to be more narrow along the $y$ axis, due to the stronger confinement in that direction. Thus, $R_<$ transitions from the bulk region to the edge earlier, giving earlier oscillations.
Another observation is that when $\omega_y$ is held constant, increasing $M$ widens the plateau. This is because the bulk density $\frac{\nu}{2\pi}=\frac{1}{2\pi M}$ decreases, giving a wider droplet as $N=300$ is held constant. 

Dashed lines in Fig.~\ref{fig:jedgenonint} indicate the opposite of the value of the quasiparticle spin, $- J_{qp}$, with $J_{qp}$ computed in Appendix~\ref{app:bulk}. This is the expected edge spin value according to the fractionalization relation anticipated in Eq.~\eqref{Eq:Fractionalization}.
It is seen that the non-interacting kernel $K_{\rm orb}$ only reproduces the expected edge spins for non-interacting ($M=1$) or circular ($\omega_y=1$) systems.
The former could be expected,
since the kernel $K_{\rm orb}$ was derived specifically for non-interacting droplets, starting from the orbitals employed to construct the Slater-determinant wavefunction. 
For a circular system, the kernel $K_{\rm orb}$, used to calculate $J^{\rm orb}_e$ as shown in Fig.~\ref{fig:jedgenonint}, coincides with the circular kernel $K_{\rm circ}$. Since $K_{\rm circ}$ has been successfully used for the circular edges of both interacting and non-interacting states, as well as for circular quasiparticles, it is not surprising that $K_{\rm orb}$ works in the circular limit.

The reason why $K_{\rm orb}$ does not work when $M>1$ and $\epsilon\neq 0$ is that the state is no longer an eigenstate of the (many-body generalization of the) single-orbital GIGR \eqref{projJ}, $\hbar\sum_{i=1}^N b_i^{\dagger}b_i$. Instead, the state becomes a linear combination of terms with different eigenvalues. This is explained further in App. \ref{app:motivation}.

This study shows that the kernel $K_{\rm orb}(x,y)$ derived from the properties of single-particle elliptic orbitals cannot be employed to define the edge spin of generic interacting quantum many-body states constructed using those orbitals. A different approach is needed. The framework for such an approach is established in the next section, together with insights on \textit{why} $K_{\rm orb}$ is inadequate for interacting states.

\section{An algebraic viewpoint on elliptic Laughlin states}\label{sec:algebraic}

In this Section, we establish an algebraic framework which explains why the edge spin kernel \eqref{ourkernel} is not applicable to arbitrary ellipse-shaped interacting many-body states. First, in Sec.~\ref{sec:derivation}, we introduce an operator deforming the well-known Hamiltonian with circular symmetry to that with elliptic symmetry. We then show in Sec.~\ref{Sec:Def:SP:Orbitals} that the same operator deforms the circular single-particle orbitals into the elliptic ones introduced in Sec.~\ref{sec:orbs}. However, in Sec.~\ref{SubSec:Deforming:Laughlin} it is seen that the elliptic Laughlin state in Eq.~\eqref{lllfun} is obtained by acting with a \textit{different} operator on the circular Laughlin state. This mismatch between orbital geometry and many-body geometry partly explains why the edge spin $J_e^{\rm orb}$, as defined using properties of the single-particle orbitals, does not generally work for the many-body state.

We remark that the framework established in this Section can be used to find an edge spin valid also for interacting systems. This is saved for Section \ref{sec:gigrmb}.

\subsection{An operator deforming circularly symmetric potentials into elliptic ones}\label{sec:derivation}

We begin by identifying an invertible operator $S$, which deforms the LLL projection of the Hamiltonian with a harmonic and rotationally symmetric potential, 
\begin{equation}
    H_{\rm circ} = H_K + V_{\rm circ},
    \quad \text{with} \;\; 
    V_{\rm circ}=\mu \frac{\omega_x\omega_y}{2}r^2,
\end{equation}
to the LLL projection of the Hamiltonian $H_K+ V_{\rm ell}$ introduced in Eq.~\eqref{pot}, which has an elliptic confining potential.
Our $S$ operator should leave $H_K$ invariant and deform the LLL projection of the potential:
\begin{equation}
    S H_K S^\dagger = H_K, 
    \qquad 
    S \tilde{V}_{\rm circ} S^\dagger = 
    \tilde{V}_{\rm ell}.
\end{equation}
We remind the reader that $\tilde{V}_{\rm circ}$ and $\tilde{V}_{\rm ell}$ are the LLL-projected versions of $V_{\rm circ}$ and $V_{\rm ell}$, respectively. 

Let us consider the LLL-projected potential $\tilde V_{\rm ell}$ in Eq.~\eqref{vlllcoord},
as well as $\tilde{V}_{\rm circ}$. The latter is
\begin{equation}\label{vcirclll}
    \tilde {V}_{\rm circ} =
    \frac{\mu\omega_x\omega_y}{2} \left[ \left(
    R^{2}_x+R_y^{2}
    \right)+\ell_B^2  \right].
\end{equation}
To stretch $\tilde V_{\rm circ}$ into $\tilde V_{\rm ell}$ without distorting the Landau levels, we introduce the operators
\begin{subequations}
\begin{align}
        \Lambda_0=&\frac{R_x^{2}+R_y^{2}}{2\ell_B^2}, \label{Eq:Lambda0:First}\\
    \Lambda_1=&\frac{R_x^{2}-R_y^{2}}{2\ell_B^2},\label{Eq:Lambda1:SP}\\
     \Lambda_2=&\frac{R_x R_y+R_y R_x }{2\ell_B^2}.
     \label{Eq:Lambda2:SP}
\end{align}
\end{subequations}
Because the guiding center coordinates obey $[R_x,R_y]=-i\ell_B^2$, the operators above have the commutation relations 
\begin{subequations}
\label{Eqs:Commutations_Lambdas}
\begin{align}
    [\Lambda_0,\Lambda_1]&=+2i\Lambda_2,\label{lambdarel0}\\
    [\Lambda_0,\Lambda_2]&=-2i\Lambda_1,\label{lambdarel1}\\
    [\Lambda_1,\Lambda_2]&=-2i\Lambda_0\label{lambdarel2}.
\end{align}
\end{subequations}

The potential $\tilde V_{\rm ell}$ in Eq.~\eqref{vlllcoord} can be written in terms of these operators:
\begin{align}\label{vlllell}
    \tilde V_{\rm ell}&=\frac{\mu\omega_x\omega_y}{2\ell_B^2}\bigg(\frac{\omega_x}{\omega_y}(\Lambda_0+\Lambda_1)+\frac{\omega_y}{\omega_x}(\Lambda_0-\Lambda_1)\bigg)+V_0\nonumber\\
    &=\frac{\mu\omega_x\omega_y}{\ell_B^2}\big(\cosh(2\lambda)\Lambda_0+\sinh(2\lambda)\Lambda_1\big)+V_0,
\end{align}
where we introduced the parameter $\lambda$, defined via 
$
e^{2\lambda}=\frac{\omega_x}{\omega_y}\label{etadef}
$; see also 
 App.~\ref{sec:lzprx}. The constant $V_0 = \frac{\mu \ell_B^2}{4} \left( \omega_x^2 + \omega_y^2 \right)$ is the one appearing in Eq.~\eqref{vlllcoord}.
Similarly, we can write
\begin{equation}\label{vcirclll2}
    \tilde{V}_{\rm circ}=\frac{\mu\omega_x\omega_y}{\ell_B^2}\Lambda_0+ V_0.
\end{equation}

Let us now introduce the desired operator $S$. Since both expressions in Eqs.~\eqref{vlllell} and \eqref{vcirclll2} lack a $\Lambda_2$ term, it is reasonable to expect the  operator $S$ to only depend on $\Lambda_2$, in light of the relations~\eqref{Eqs:Commutations_Lambdas}. Indeed, taking the unitary operator
\begin{equation}\label{squeeze}
    S=e^{-i\lambda \Lambda_2}
\end{equation}
and using Campbell's lemma \cite{campbell1896law},
\begin{equation}\label{campbell}
    e^{A}Be^{-A}=B+[A,B]+\frac{1}{2!}[A,[A,B]]+...\;,
\end{equation}
we find that
\begin{align}\label{L0squeeze}
    S\Lambda_0S^{\dagger}=\cosh(2\lambda)\Lambda_0+\sinh(2\lambda)\Lambda_1.
\end{align}
This implies that
\begin{align}\label{vcircsqueeze}
    S\tilde{V}_{\rm circ}S^{\dagger}&=\frac{\mu\omega_x\omega_y}{\ell_B^2}\big(\cosh(2\lambda)\Lambda_0+\sinh(2\lambda)\Lambda_1\big)+V_0
    \nonumber\\
    &=\tilde {V}_{\rm ell}.
\end{align}
Since $S$ only depends on the guiding center operators, which commute with the mechanical momenta $\pi_x$ and $\pi_y$, the transformation $S$ leaves $H_K$ invariant.
We can thus conclude that we found the operator we were looking for.

There is a corollary of the above worth mentioning. Eq.~\eqref{L0squeeze} with $e^{2\lambda}=\frac{\omega_x}{\omega_y}$ and Eqs.~\eqref{Eq:Lambda0:First} and \eqref{Eq:Lambda1:SP} imply that 
\begin{align}\label{trafogigr}
    S\bigg(\frac{R_x^2+R_y^2}{2\ell_B^2}\bigg)S^{\dagger}&=\frac{\omega_x}{2\omega_y}(\Lambda_0+\Lambda_1)+\frac{\omega_y}{2\omega_x}(\Lambda_0-\Lambda_1)\nonumber\\
    &=\frac{1}{2\ell_B^2}\bigg(\frac{\omega_x}{\omega_y}R_x^2+\frac{\omega_y}{\omega_x}R_y^2\bigg).
\end{align}
From Eqs. \eqref{jcirc}, \eqref{jell} and \eqref{trafogigr}, it follows that
\begin{equation}
    SJ_{\rm circ}S^{\dagger} = J_{\rm orb}.
\end{equation}
Above, we have used that since $S$ acts trivially on the momenta $\pi_x$ and $\pi_y$, the momentum-dependent term in Eq. \eqref{jcirc} transforms trivially under $S$. We conclude that the $S$ operator does not only map the two potentials into each other, but also relates the corresponding GIGRs.

We remark that closely related transformations have previously been applied to metrics in QH systems \cite{park2014dipole,yuzhu2023geometric}. For relevant mathematical background, we refer the reader to Ref. \cite{puri2001mathematical}. Additionally, Ref. \cite{oblak2023adiabatic} performs a detailed study of deformations in QH systems. The authors investigate generalizations of magnetic translations, and since our $S$ operator is constructed using guiding center operators, one can speculate that $S$ fits within their framework.

\subsection{Deforming the single-particle orbitals}\label{Sec:Def:SP:Orbitals}

The operator $S=e^{-i\lambda \Lambda_2}$ from Eq.~\eqref{squeeze} can be used to revisit the discussion on the properties of the eigenvectors of $\tilde V_{\rm ell}$ in the LLL.
Starting from the well-known fact that the eigenvectors of any rotationally invariant potential projected to the LLL are of the form $\phi_m \propto z^m e^{-|z|^2 / 4 \ell_B^2}$, and that $\tilde V_{\rm circ} \phi_m = E_m \phi_m$, we can find the eigenvectors of $\tilde V_{\rm ell}$ with the help of the $S$ operator.
Observing that
$    
S \tilde{V}_{\rm circ} S^\dagger S \phi_m
= E_m S \phi_m,
$
we conclude that
\begin{equation}
    \tilde V_{\rm ell} \, S \phi_m = 
    E_m \, S \phi_m.
\end{equation}
The states $S \phi_m$ therefore have the same eigenvalues for the LLL-projected potential energy as our states $\psi_m$ introduced in Sec.~\ref{sec:orbs} (c.f. Eq.~\eqref{poteigen}). Because the potential energy is non-degenerate, $S\phi_m$ coincides with $\psi_m$, up to a possible proportionality factor. Since $S$ is unitary, using the properly normalized $\phi_m$ means that such a prefactor has to have modulus $1$.

Interestingly, it can also be explicitly verified that our $S$ operator in Eq.~\eqref{squeeze} takes the circular orbitals $\phi_m$ to the ellipse-shaped ones, i.e., $S\phi_m = \psi_m$. In other words, the two are not just proportional, as argued earlier, but equal. We provide the detailed calculation in App.~\ref{app:orbdefdetails}. For our current purposes, we simply note that the operator $S$ can deform both the potential and the orbitals for the rotationally symmetric system to those of the ellipse-shaped one. 

\subsection{Deforming the interacting Laughlin states}
\label{SubSec:Deforming:Laughlin}
Let us now turn to the many-body state \eqref{lllfun}. From the previous section, it is natural to try to obtain this state by deforming the standard (rotationally symmetric) Laughlin state, $\Psi_{\nu=\frac{1}{M}}^{\epsilon=0}$, using a many-body version of $S$. For $M=1$, this will be shown to produce our state \eqref{iqhfun}. However, when $M>1$, this deformation of $\Psi_{\nu=\frac{1}{M}}^{\epsilon=0}$ will be seen to give a state which differs from \eqref{lllfun} and which is unsuitable on physical grounds. Thus, we reject this state in favor of our state \eqref{lllfun}. In order to obtain the latter from the circular Laughlin state, a different deformation needs to be introduced. The discrepancy between the two deformations explains why the single-particle kernel is valid for $J_e$ when $M=1$, but not when $M>1$. All of this is described in greater detail below. 

\subsubsection{Orbital deformation of the circular Laughlin states}
Let us now deform the non-interacting circular Laughlin state at $\nu = 1$, $\Psi_{\nu=1}^{\epsilon=0}$. To do so, we introduce the many-body version of $S$,
\begin{equation}\label{smb}
    S^{\rm (MB)}=e^{-i\lambda \Lambda_2^{\rm (MB)}},
\end{equation}
where the many-body generalizations of Eqs.~\eqref{Eq:Lambda0:First}-\eqref{Eq:Lambda2:SP} are given by
\begin{subequations}
\begin{align}
    \Lambda_0^{\rm (MB)} &= \sum_{a=1}^N \Lambda_0^{(a)} =\sum_{a=1}^N
    \frac{(R_x^a)^2 + (R_y^a)^2}{2 \ell_B^2},\label{lambda0mb}\\
    \Lambda_1^{\rm (MB)} &= \sum_{a=1}^N \Lambda_1^{(a)}= \sum_{a=1}^N
    \frac{(R_x^a)^2 - (R_y^a)^2}{2 \ell_B^2},\label{lambda1mb}\\
    \Lambda_2^{\rm (MB)} &= \sum_{a=1}^N \Lambda_2^{(a)} = \sum_{a=1}^N
    \frac{R_x^a R_y^a + R_y^a R_x^a}{2 \ell_B^2}\label{lambda2mb}.
\end{align}
\end{subequations}
Since operators acting on different particle coordinates commute, one trivially has 
\begin{equation}
    S^{\rm (MB)}=\prod_{a=1}^N e^{-i\lambda \Lambda_2^{(a)}}\equiv \prod_{a=1}^N S^{(a)},
\end{equation}
where each factor is a one-body $S$ operator $S^{(a)}\equiv e^{-i\lambda \Lambda_2^{(a)}}$ acting on particle $a$. We also know that the polynomial part of the circular Laughlin state for $M\geq 1$ can be written as a Slater determinant raised to a power:
\begin{align}
    \Psi_{\nu=\frac{1}{M}}^{\epsilon=0} &= \big({\rm sl}_\lambda \big)^Me^{-\frac{1}{4\ell_B^2}\sum_j |z_j|^2}\nonumber\\&=\bigg(\mathcal{A} \Bigl[
 \phi'_{\lambda_1}(z_1)
\phi'_{\lambda_2}(z_2)\cdots
\phi'_{\lambda_N}(z_N)
\Bigr]\bigg)^Me^{-\frac{1}{4\ell_B^2}\sum_j |z_j|^2}.
\end{align}
Above, we use $\phi_m'$ to denote the polynomial part of the circular orbital $\phi_m$. In other words, $\phi'_m(z)\propto z^m$, excluding the Gaussian, and $\phi_m(z,\bar{z})=\phi'_m(z)e^{-\frac{1}{4\ell_B^2}|z|^2}$. The state above simply deforms as
\begin{align}\label{mbtrafo}
    S^{\rm (MB)}\Psi_{\nu=\frac{1}{M}}^{\epsilon=0} &=\prod_{a=1}^N S^{(a)}\bigg(\mathcal{A} \Bigl[
 \phi'_{\lambda_1}(z_1)\cdots
\phi'_{\lambda_N}(z_N)
\Bigr]\bigg)^M\nonumber\\
&\times e^{-\frac{1}{4\ell_B^2}\sum_j |z_j|^2},
\end{align}
where the $S^{(a)}$ are understood to act on everything on their right, including the Gaussians. For $M=1$, every term in the anti-symmetrized expression is a product of the (polynomial parts of the) original circular orbitals, and each pair of particle coordinates $(z_a,\bar{z}_{a})$ occurs only within one orbital per term. Thus,
\begin{align}
    S^{\rm (MB)}\Psi_{\nu=1}^{\epsilon=0}&=\mathcal{A} \Bigl[
 S^{(1)}\phi_{\lambda_1}(z_1)\cdots
S^{(N)}\phi_{\lambda_N}(z_N)
\Bigr]\nonumber\\&=\mathcal{A} \Bigl[
 \psi_{\lambda_1}(z_1)\cdots
\psi_{\lambda_N}(z_N)
\Bigr],
\label{eq:SmbonM1}
\end{align}
where we have expressed the RHS in terms of the full orbital wavefunctions and used the deformation of the single-particle orbitals from Sec.~\ref{Sec:Def:SP:Orbitals}. We recognize the last line as the Slater determinant of our orbitals $\psi_m$ from~\eqref{orbs}. From Eq.~\eqref{iqhfun}, this is exactly our elliptic integer quantum Hall state $\Psi_{\nu=1}$, so we may conclude that
\begin{equation}
    S^{\rm (MB)}\Psi_{\nu=1}^{\epsilon=0} = \Psi_{\nu=1}.
\end{equation}

Let us now consider $M>1$ in Eq.~\eqref{mbtrafo}, and show that this does \textit{not} give our state $\Psi_{\nu=\frac{1}{M}}.$ In this case, we cannot continue as in Eq.~\eqref{eq:SmbonM1}. For $M>1$, raising the antisymmetrized expression in Eq.~\eqref{mbtrafo} to the power $M$ and then deforming the orbitals generally gives a different result from when one performs the operations in the other order. This is due to the structure of the Hermite polynomials. 
Therefore, one cannot make the transition from Eq.~\eqref{mbtrafo} to the $M>1$ analogue of the first line of Eq.~\eqref{eq:SmbonM1}. This step is only valid for $M=1$, where the exponentiation is trivial.

To find the RHS of Eq.~\eqref{mbtrafo} for $M>1$, we reason as follows.
The antisymmetrization $\mathcal{A}[...]$ in the RHS of Eq.~\eqref{mbtrafo} gives a sum of terms where each particle coordinate $z_i$ occurs in exactly one orbital per term. The exponent $M>1$ then leads to a sum of terms, where in each term, a given particle coordinate $z_i$ occurs in $M$ factors (including the possible factor $z_i^0=1$). Each such term is of the form $\prod_{j=1}^M\phi'_{\lambda_{i_j}}(z_i)$, where the different parts $\lambda_{i_1},...,\lambda_{i_M}$ may or may not differ. Due to the form $\phi'_{\lambda_j}(z_i)\propto z_i^{\lambda_j}$ of the (polynomial parts of the) circular orbitals, it follows that
\begin{equation}
    \prod_{j=1}^M\phi'_{\lambda_{i_j}}(z_i) \,\propto\, z_i^{\sum_{j=1}^M\lambda_{i_j}} \,\propto\, \phi'_{\sum_{j=1}^M\lambda_{i_j}}(z_i),
\end{equation}
which together with its $e^{-\frac{1}{4\ell_B^2}|z_i|^2}$ factor in Eq.~\eqref{mbtrafo} is still one of the circular LLL orbitals. Let us define $|\lambda_i|\equiv\sum_{j=1}^M\lambda_{i_j}$ (which should not be mistaken for the absolute value of the $\lambda_i$ in Eq.
\eqref{mbtrafo}). Then the $M>1$ case of Eq.~\eqref{mbtrafo} 
becomes a linear combination of various terms of the form 
\begin{align}
    \prod_{a=1}^NS^{(a)}\phi_{|\lambda_1|}(z_1)\cdots \phi_{|\lambda_N|}(z_N) = \psi_{|\lambda_1|}(z_1)\cdots \psi_{|\lambda_N|}(z_N).
\end{align}
This linear combination is in general not equal to our state~\eqref{lllfun}, which instead has terms $\phi_{|\lambda_1|}(z_1)\cdots \phi_{|\lambda_N|}(z_N)e^{\frac{\epsilon}{4\ell_B^2}\sum_s z_s^2}$, where the polynomial part is different. For instance, the RHS of Eq.~\eqref{mbtrafo} for $N=2,M=2$ becomes
\begin{widetext}
\begin{align}
    S^{(1)}S^{(2)}(z_1-z_2)^2e^{-\frac{1}{4\ell_B^2}\sum_{j=1}^2 |z_j|^2} &= 
    S^{(1)}S^{(2)}(z_1^2z_2^0-2z^1_1z^1_2+z_1^0z_2^2)e^{-\frac{1}{4\ell_B^2}\sum_{j=1}^2 |z_j|^2}\nonumber\\
    &= \psi_2(z_1)\psi_0(z_2)-2\psi_1(z_1)\psi_1(z_2)+\psi_0(z_1)\psi_2(z_2).
\end{align}
Using the expressions~\eqref{orbs} and~\eqref{coeffs}, where the latter implies $C_1=\sqrt{\frac{\epsilon}{2}}C_0$ and $C_2=\frac{\epsilon}{2\sqrt{2}}C_0$, we obtain that
\begin{align}\label{2part}
    S^{(1)}S^{(2)}(z_1-z_2)^2e^{-\frac{1}{4\ell_B^2}\sum_{j=1}^2 |z_j|^2}
    &= \sqrt{2}\epsilon C_0^2\bigg[\zeta_1^2+\zeta_2^2-2\sqrt{2}\zeta_1\zeta_2-1\bigg] e^{\frac{\epsilon}{4\ell_B^2}\sum_{j=1}z_j^2}e^{-\frac{1}{4\ell_B^2}\sum_{j=1}^2 |z_j|^2},
\end{align}
\end{widetext}
where $\zeta_i = \frac{z_i}{\ell_B} \sqrt{\frac{\omega_x \omega_y}{\omega_y^2 - \omega_x^2}}$. It is clear that we do not obtain the two-particle case of $\Psi_{\nu=\frac{1}{2}}$ in Eq.~\eqref{qhfun}. Similar arguments for other $N$ and $M>1$ imply that deforming the $\Psi_{\nu=\frac{1}{M}}^{\epsilon=0}$ using $S^{\rm (MB)}$ does not give~\eqref{lllfun}.

There is another, more serious problem with the state in Eq. \eqref{2part}: the wavefunction does not vanish when $\zeta_1=\zeta_2$ (or equivalently, $z_1=z_2$). Consequently, a short-range interaction $V_{\rm int}=\delta(z_1-z_2)$ develops a non-zero expectation value $\langle V_{\rm int}\rangle\neq 0$. This problem is not unique for $N=2$, and means that if we wish to find low-energy states, $S^{\rm (MB)}\Psi_{\nu=\frac{1}{M}}^{\epsilon=0}$ is not a suitable candidate. Because our state $\Psi_{\nu=\frac{1}{M}}$ does not have this problem, we study $\Psi_{\nu=\frac{1}{M}}$ instead of $S^{\rm (MB)}\Psi_{\nu=\frac{1}{M}}^{\epsilon=0}$.

\subsubsection{A many-body deformation valid for interacting states}
It has been shown that $S^{\rm (MB)}\Psi_{\nu=\frac{1}{M}}^{\epsilon=0}$ is not generally a low-energy state for $M>1$ given that the particles have short-range interactions. We instead wish to deform the circular Laughlin state $\Psi_{\nu=\frac{1}{M}}^{\epsilon=0}$ into our (physically reasonable) state $\Psi_{\nu=\frac{1}{M}}$. Thus, we require a deformation different from $S^{\rm (MB)}$. For this purpose, we first define the auxiliary operators
\begin{equation}\label{lambdapm}
    \Lambda_{\pm}=\frac{\Lambda_1\pm i\Lambda_2}{\sqrt{2}},
\end{equation}
where $\Lambda_1$ and $\Lambda_2$ are given by Eqs.~\eqref{lambda1mb} and~\eqref{lambda2mb}.
Note that in order to avoid unnecessarily heavy notation, we suppress the apex $\rm(MB)$ to indicate many-body operators. This will be obvious from the context, since from here on we will be dealing exclusively with the many-body problem.
Using the commutation relations for $\{ \Lambda_0, \Lambda_1, \Lambda_2 \}$ in Eq.~\eqref{Eqs:Commutations_Lambdas} for each particle index $a$ in Eqs.~\eqref{lambda0mb}-\eqref{lambda2mb}, it follows that
\begin{equation}\label{commrelspm}
    [\Lambda_0,\Lambda_{\pm}]=\pm 2\Lambda_{\pm},\quad [\Lambda_+,\Lambda_-]=-2\Lambda_0.
\end{equation}

We want to use the operators $\{\Lambda_0,\Lambda_+,\Lambda_-\}$ in order to construct an operator that transforms $\Psi^{\epsilon=0}_{\nu=\frac{1}{M}}$ into $\Psi_{\nu=\frac{1}{M}}$.
These wavefunctions lie in the lowest Landau level, which means that we should project $\{\Lambda_0,\Lambda_+,\Lambda_-\}$ onto the lowest Landau level. We use the Girvin-Jach (or Bargmann) picture~\cite{girvin1984formalism,bargmann1962representations}, and denote the resulting operators by $\{ \Upsilon_0, \Upsilon_+, \Upsilon_- \}$. It is implicitly understood that derivatives appearing in these operators {\em do not} act on the Gaussian factors $e^{-|z|^2/(4\ell_B^2)}$ present in the wavefunctions, as prescribed by the Girvin-Jach procedure. Because the derivatives in the operators $\{\Lambda_0,\Lambda_+,\Lambda_-\}$ do act on the Gaussian factors $e^{-|z|^2/(4\ell_B^2)}$, one needs to be careful when projecting $\partial_z$ and $\partial_{\bar{z}}$ onto the lowest Landau level. One obtains the following operators (which indeed satisfy the commutators~\eqref{commrelspm}):
\begin{subequations}
\label{rep}
\begin{align}
    \Upsilon_0&=\sum_{a=1}^Nz_a\frac{\partial}{\partial z_a}+\frac{N}{2}=L_0+\frac{N}{2},\\
    \Upsilon_+&=\frac{1}{2\sqrt{2}}\sum_{a=1}^N\frac{z_a^2}{\ell_B^2},\qquad
    \Upsilon_-=\sqrt{2}\ell_B^2\sum_{a=1}^N\frac{\partial^2}{\partial z_a^2}.
\end{align}
\end{subequations}
Using this representation, the elliptic Laughlin state we have introduced in Eq.~\eqref{lllfun} can also be obtained by deforming the circular Laughlin state. However, the deformation used is different from that in Eq.~\eqref{mbtrafo}: 
\begin{equation}\label{deformlaughl}
    \Psi_{\nu = \frac 1M} \propto e^{s_+ \Upsilon_+} \Psi^{\epsilon=0}_{\nu=\frac{1}{M}},
    \qquad s_+ = \frac{\epsilon}{\sqrt 2}.
\end{equation}
This transformation is rather trivial, since the operator simply acts like a multiplicative factor, giving our state~\eqref{lllfun} the proper anisotropic Gaussian. Less trivial properties of the operators in Eq.~\eqref{rep} will be used in Sec.~\ref{Sec:Laugh:GoodGenerator}. 

That the transformations in Eqs.~\eqref{mbtrafo} and~\eqref{deformlaughl} differ allows us to understand why $J_e^{\rm orb}$ gives the expected spin fractionalization when $M=1$ but not when $M>1$ unless the droplet is circular (see Sec.~\ref{sec:edgenonint} and Fig.~\ref{fig:jedgenonint}). For $M=1$, our many-body state~\eqref{lllfun} can be obtained from the circular case by simply using the same transformation as for the single-particle orbitals, once for each particle. The orbitals and the many-body state have the same geometric properties, and it is acceptable to use a kernel derived using properties of the elliptic single-particle orbitals (as was done with $K_{\rm orb}$ in Sec.~\ref{SubSec:Integral:Kernel}) also for the many-body case.

By contrast, when $M>1$, the orbital transformation $S^{\rm (MB)}$ no longer gives our state~\eqref{lllfun}. Our transformation $e^{s_+\Upsilon_+}$ gives a different geometry, and there is no longer any reason for $K_{\rm orb}$ to be valid.

For completeness, we also remark that in the circular limit, $\lambda=\frac{1}{2}\ln(\omega_x/\omega_y)=0$ and $s_+=\frac{\epsilon}{\sqrt{2}}=0$. This gives $S^{\rm (MB)}=I$ and $e^{s_+\Upsilon_+}=I$. Therefore, our state $\Psi_{\nu=\frac{1}{M}}$ is trivially equal to $S^{\rm (MB)}\Psi_{\nu=\frac{1}{M}}^{\epsilon=0}$ in that case, and the single-particle kernel $K_{\rm orb}$ should give the expected edge spin regardless of $M$. That this holds is apparent from Fig. \ref{fig:jedgenonint}. 

Based on the previous discussion, it is natural to instead try to derive an edge spin kernel for $\Psi_{\nu=\frac{1}{M}}$ based on the transformation \eqref{deformlaughl} of the many-body state. This will be done in Sec. \ref{sec:gigrmb}. We will find that the operators in Eq.~\eqref{rep} can be used to deform the regular LLL angular momentum operator $L_z=\hbar z\partial_z$ (in the Girvin-Jach picture) to a GIGR suitable for generic ellipse-shaped states. 

\section{Generator of rotations for the elliptic Laughlin states}\label{sec:gigrmb}

We now proceed to find a suitable generator of rotations that is applicable to the interacting ellipse-shaped Laughlin states that we introduced in Sec.~\ref{sec:many-body}. 
In Sec.~\ref{Sec:Laugh:GoodGenerator}, we find an integral kernel representation of this generator by deforming the circular angular momentum operator 
using the operator $e^{s_+ \Upsilon_+}$ introduced at the end of the previous section. We then give an operator representation. As will be shown, both representations are related to an emergent ellipticity of the density profile. This ellipticity
differs from that of the elliptic confining potential~\eqref{pot}, which gives rise to the multiplicative pre-factor $e^{\frac{\epsilon}{4 \ell_B^2}\sum_s z_s^2}$ in the elliptic Laughlin states
(though we should mention that we do not know if these states are the ground states of the elliptic potential in the presence of a suitable repulsive interaction). 
We interpret this difference in geometry in terms of an emergent metric.
In Sec.~\ref{sec:edgeint}, we numerically demonstrate that the generator of rotations derived in Sec. \ref{Sec:Laugh:GoodGenerator} can be used to define a fractionally quantized edge spin. This is one of the key results of the present work.

\subsection{The generator of rotations of the elliptic Laughlin state}
\label{Sec:Laugh:GoodGenerator}
We now wish to find a GIGR suitable for the state~\eqref{lllfun} by deforming the generator of rotations for a circular droplet, which is the usual angular momentum operator $L_z$. Since the many-body state is obtained from the circular Laughlin state using the transformation in Eq.~\eqref{deformlaughl}, we use the same transformation for $L_z$. 

What follows relies on the operators introduced in Sec.~\ref{sec:algebraic}. Eq.~\eqref{rep} can be used to write $L_z$ as follows:
\begin{equation}\label{Lzop}
    L_z=\sum_{a=1}^N z_a\frac{\partial}{\partial z_a}=\Upsilon_0-\frac{N}{2},
\end{equation}
where we set $\hbar=1$.

We now wish to deform $L_z$ to make it suitable for an elliptic Laughlin state:
\begin{equation}\label{deformLzOp}
    L_z\rightarrow L_z' = e^{s_+\Upsilon_+}L_ze^{-s_+\Upsilon_+},
\end{equation}
where we call the transformed operator $L_z'$. We point out that the above similarity transformation uses $\Upsilon_+$ only, with no reference to $\Upsilon_+^{\dagger}=\Upsilon_-$. From Eqs.~\eqref{commrelspm} and \eqref{campbell}, it follows that
\begin{equation}
    e^{s_+\Upsilon_+}\Upsilon_0 e^{-s_+\Upsilon_+}=\Upsilon_0-2s_+\Upsilon_+.
\end{equation}
Therefore, our transformed angular momentum operator becomes
\begin{equation}\label{squeezedlz}
    L_z'=\Upsilon_0-2s_+\Upsilon_+-\frac{N}{2}
    =L_z-\frac{\epsilon}{2}\sum_{a=1}^N\frac{z_a^2}{\ell_B^2},
\end{equation}
using our expression for $\Upsilon_+$ as well as~\eqref{deformlaughl} in the last step. Although $L_z$ (as given in Eq.~\eqref{Lzop}) is Hermitian in the Bargmann picture, the last term in Eq.~\eqref{squeezedlz} is not, unless $\epsilon=0$. Therefore, our operator $L_z'$ is in general non-Hermitian.

A merit of our deformed angular momentum operator $L_z'$ is that it has our ellipse-shaped state \eqref{lllfun} as an eigenstate, just like the circular Laughlin state $\Psi_{\nu=\frac{1}{M}}^{\epsilon=0}$ is an eigenstate of $L_z$. Furthermore, the operator \eqref{squeezedlz} has the same eigenvalue when acting on \eqref{lllfun} as $L_z$ has when acting on $\Psi_{\nu=\frac{1}{M}}^{\epsilon=0}$, which gives the eigenvalue of \eqref{squeezedlz} the natural interpretation of a (generalized) angular momentum.

The above can be shown as follows. 
We know that the circular Laughlin state obeys
\begin{equation}\label{eigvalcircLaughlin}
    L_z\Psi_{\nu=\frac{1}{M}}^{\epsilon=0}=\ell_0\Psi_{\nu=\frac{1}{M}}^{\epsilon=0},
\end{equation}
where $\ell_0=\frac{M}{2}N(N-1)$ is the angular momentum eigenvalue for an $N$-particle Laughlin system with no quasiholes and the filling fraction $\nu=\frac{1}{M}$.
The above implies that
\begin{equation}
    e^{s_+\Upsilon_+}L_ze^{-s_+\Upsilon_+}e^{s_+\Upsilon_+}\Psi_{\nu=\frac{1}{M}}^{\epsilon=0}=\ell_0e^{s_+\Upsilon_+}\Psi_{\nu=\frac{1}{M}}^{\epsilon=0}.
\end{equation}
Using the deformation of the circular Laughlin state in Eq. \eqref{deformlaughl}, and the deformation of $L_z$ in \eqref{deformLzOp}, we have that
\begin{equation}\label{eig0}
    L_z'\Psi_{\nu=\frac{1}{M}}=\ell_0\Psi_{\nu=\frac{1}{M}}.
\end{equation}
Similar manipulations can also be used for the state \eqref{qhfun}, if the quasihole is at the origin (otherwise the circular Laughlin state is not an $L_z$ eigenstate). In that case, Eq. \eqref{eigvalcircLaughlin} is replaced by
\begin{equation}
    L_z\Psi_{\nu=\frac{1}{M},\mathrm{qh}}^{\epsilon=0}=(\ell_0+N)\Psi_{\nu=\frac{1}{M},\mathrm{qh}}^{\epsilon=0}.
\end{equation}
Using the same deformations as for the state without quasiholes then gives
\begin{equation}\label{eig1}
    L_z'\Psi_{\nu=\frac{1}{M},\mathrm{qh}}=(\ell_0+N)\Psi_{\nu=\frac{1}{M},\mathrm{qh}}.
\end{equation}
The relations in Eqs. \eqref{eig0} and \eqref{eig1} are entirely analogous to those for the circular Laughlin state. Thus, we have established that the operator $L_z'$ behaves as expected for an angular momentum.

\subsubsection{The kernel representation of the generator of rotations}
Having obtained our generator of rotations $L_z'$, let us now represent it using an integration kernel. 
We would like to represent the  relations \eqref{eig0} and \eqref{eig1} using integrals of the form
\begin{equation}\label{jgenericmb}
    J_{\rm tot,i}=\int K_{\rm int}(r,\phi)\rho_i(r,\phi) rdrd\phi,
\end{equation}
where $K_{\rm int}$ is some integration kernel applicable also to interacting and non-circular systems. Here, $\rho_i$ can be taken as the density $\rho_0$ for a droplet without any quasiholes, corresponding to Eq. \eqref{eig0}. This should give the total angular momentum $J_{\rm tot, 0}=\ell_0$, as in Eq. \eqref{eig0}. Alternatively, one can take $\rho_i=\rho_{qh}$, which is the density of a droplet with a quasihole at the origin. This should yield $J_{\rm tot, qh}=\ell_0+N$ as in Eq. \eqref{eig1}. 

The densities depend on $|\Psi_{\nu=\frac{1}{M}}|^2$ and $|\Psi_{\nu=\frac{1}{M},\mathrm{qh}}|^2$, respectively, with all coordinate pairs $(x_{a},y_{a})$ except one integrated out. These densities can be computed numerically using e.g. Metropolis sampling or matrix product states (MPS), where the MPS computation naturally leads to a density depending only on two spatial coordinates. From now on, we will use the MPS-appropriate notation, for consistency with previous work \cite{fagerlund2025spin}. This means dropping the particle index $a$, with the understanding that while $L_z'$ depends on only one pair of coordinates $(x,y)$, it still acts on the density of a many-body state.  

Let us now find our kernel $K_{\rm int}$. We know that the angular momentum operator $L_z$ in Eq. \eqref{squeezedlz} can be represented as the integration kernel $\frac{r^2}{2\ell_B^2}-1$ when acting on the many-body density \cite{comparin2022spin,nardin2023ssr}. The second term in the RHS of Eq.~\eqref{squeezedlz} (with the sum over the particle index $a$ removed) is proportional to $z^2=x^2+2ixy-y^2$, and for an ellipse-shaped droplet, the symmetry of the density under $x\rightarrow -x$ and $y\rightarrow -y$ means that the imaginary term will integrate to $0$. Hence, it can be omitted, giving the real-valued kernel
\begin{align}\label{intkernel}
    K_{\rm int}&=\hbar\bigg(\frac{r^2}{2\ell_B^2}-1-\frac{\epsilon}{2\ell_B^2}(x^2-y^2)\bigg)\nonumber\\
    &=\hbar\bigg(\frac{r^2\big(1-\epsilon\cos(2\phi)\big)}{2\ell_B^2}-1\bigg),
\end{align}
where we have restored $\hbar$. This is the integral kernel that we propose for defining a quantized edge spin for the elliptic Laughlin state. 
It represents a gauge invariant generator of rotations for the system, as it applies no matter the gauge employed to write the state.
As will be seen in Sec.~\ref{sec:edgeint}, the quantized edge spin will be defined as a difference of two integrals of the form \eqref{jgenericmb}, with the kernel given by Eq. \eqref{intkernel}, and with certain integration limits imposed. Before investigating this further, however, we wish to discuss the geometry associated with $K_{\rm int}$ in more detail.

\begin{figure}
    \centering
    \includegraphics[width=\linewidth]{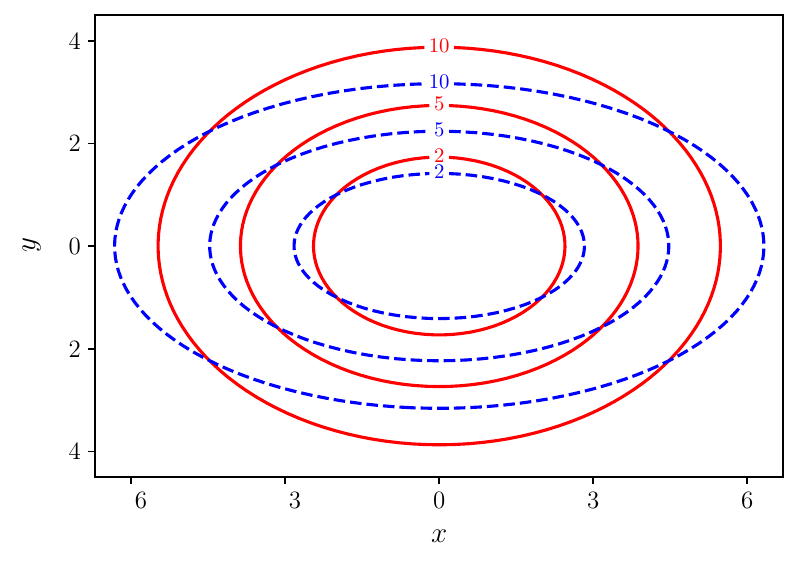}
    \caption{A few level curves of the potential in Eq.~\eqref{pot} (blue, dashed) and the position-dependent part of the kernel in Eq.~\eqref{intkernel} (red, plain). Both sets of curves are drawn for $\omega_x=1$ and $\omega_y=2$, and we have set $\mu=\frac{1}{2}$ for the potential, as well as $\hbar=1$ and $\ell_B=1$. The number on each level curve shows the value of the function being plotted, i.e., of $V_{\rm ell}$ or $K_{\rm int} +1$.}
    \label{fig:levelcurves}
\end{figure}

We note that a level curve of the kernel~\eqref{intkernel} describes an ellipse $\frac{x^2}{a_k^2}+\frac{y^2}{b_k^2}=1$, where the semi-axes $a_k$ and $b_k$ obey $\frac{1}{a_k^2}\propto 1-\epsilon = \frac{2\omega_x}{\omega_x+\omega_y}$ and $\frac{1}{b_k^2}\propto 1+\epsilon =\frac{2\omega_y}{\omega_x+\omega_y}$, with the same proportionality factor for both. This gives $\frac{a_k}{b_k}=\sqrt{\frac{\omega_y}{\omega_x}}$.
 By examining Metropolis sampled densities (see App.~\ref{app:bulk}), one finds that the axis lengths $a_{\rho}$ and $b_{\rho}$ of the \textit{droplet} scale as
\begin{equation}
    a_{\rho}\propto \sqrt{\frac{\omega_y}{\omega_x}},\quad b_{\rho}\propto \sqrt{\frac{\omega_x}{\omega_y}},
\end{equation}
giving an area $A=\pi a_{\rho} b_{\rho}$ which is independent of the ratio of the axes, as it should be. It follows that $\frac{a_{\rho}}{b_{\rho}}\propto\frac{\omega_y}{\omega_x}$, and by considering the special case of a circular droplet, one realizes that the proportionality constant is $1$, i.e., $\frac{a_{\rho}}{b_{\rho}}=\frac{\omega_y}{\omega_x}$. This is seen also in Fig. \ref{fig:axes}. It follows that while the level curves of \eqref{intkernel} for a given $\epsilon$ describe ellipses, they do not have the same shape as the elliptic density of the droplet for the same $\epsilon$. We show this in Fig.~\ref{fig:levelcurves}, where the level curves of the potential in Eq.~\eqref{pot} are plotted against the same level curves of the position-dependent part of $K_{\rm int}$, i.e., those of $K_{\rm int}+1$ when $\hbar=1$. 

The above implies that 
the appropriate edge spin kernel does not necessarily follow the geometry of the
droplet.
However, for the straight edge of the MPS cylinder studied in Ref.~\cite{fagerlund2025spin}, 
and for a circular droplet, the  kernel and droplet geometries coincide. 

A similar scaling analysis for the non-interacting kernel \eqref{ourkernel} gives $\frac{a_k}{b_k}\propto \frac{\omega_y}{\omega_x}$. Thus, even if the interacting kernel does not follow the shape of the droplet, the non-interacting kernel \textit{does}. The principle that the kernel \textit{“follows the droplet geometry"} therefore only holds for systems which are circular or non-interacting (or both). The failure of this principle can be viewed as an emergent kernel geometry induced by interactions away from the circular limit, see Sec.~\ref{subsec:metric}.

We now proceed to give the gauge-invariant generator of rotations corresponding to the kernel \eqref{intkernel}, written in a fully gauge invariant way. It reads 
\begin{equation}\label{jint}
    J_{\rm int}
    =\frac{\hbar}{2\ell_B^2}\bigg(\sqrt{\frac{\omega_x}{\omega_y}}R^2_x+\sqrt{\frac{\omega_y}{\omega_x}}R^2_y\bigg)-\frac{\ell_B^2}{2\hbar}\big(\pi_x^2+\pi_y^2\big).
\end{equation}
With \eqref{jint}, we can also find the commutators replacing \eqref{commsell}, which become
\begin{align}\label{commsint}
    [J_{\rm int},\pi_x]&=i\hbar\pi_y,\;\;\qquad\quad [J_{\rm int},\pi_y]=-i\hbar\pi_x, \nonumber\\  [J_{\rm int},R_x]&=i\hbar\sqrt{\frac{\omega_y}{\omega_x}}R_y,\quad [J_{\rm int},R_y]=-i\hbar\sqrt{\frac{\omega_x}{\omega_y}}R_x.
\end{align}
Even if the commutation relations manifestly depend on the shape of the potential (i.e.~the quotient $\omega_y/\omega_x$), they do so differently than in \eqref{commsell}. We view this difference as being due to the presence of interactions, and due to the different deformation ($e^{s_+\Upsilon_+}$ rather than $S^{\rm (MB)}$) required for an interacting state.

\subsubsection{Interpreting the generator of rotations via a metric}\label{subsec:metric}

Another perspective on the kernel in Eq.~\eqref{intkernel} is the following. Since it is a quadratic form in the spatial coordinates, it may be written as a “distance" using a metric:
\begin{align}
    \frac{K_{\rm int}}{\hbar}
    &=
    \frac{\omega_x}{(\omega_x+\omega_y)\ell_B^2}x^2+\frac{\omega_y}{(\omega_x+\omega_y)\ell_B^2}y^2-1\nonumber\\
     &=g_{\rm int}^{ij}x_ix_j-1,
\end{align}
with $x_1=x,x_2=y$ and 
\begin{equation}\label{gint}
    g_{\rm int}=\frac{1}{(\omega_x+\omega_y)\ell_B^2}\begin{bmatrix}
        \omega_x&0\\
        0&\omega_y
    \end{bmatrix}.
\end{equation}
The idea is that this metric represents an emergent geometry, which differs from that of the QH droplet.
Note that the result depends on $\omega_y/\omega_x$ but not on $M$. 
The non-interacting kernel in Eq.~\eqref{ourkernel}  instead corresponds to
\begin{equation}\label{gnonint}
    g_{\rm orb}=\frac{1}{2\ell_B^2}\begin{bmatrix}
        \frac{\omega_x}{\omega_y}&0\\
        0&\frac{\omega_y}{\omega_x}
    \end{bmatrix}.
\end{equation}
Consequently, there is a modified kernel, and a modified metric, arising due to the presence of interactions. 

We note that even though both $K_{\rm int}$ and $K_{\rm orb}$ are valid in the non-interacting limit $M\rightarrow 1$, they do not become equal in this limit, since neither depends on $M$. However, in the circular limit, both kernels coincide, as do the GIGR \eqref{jell} and \eqref{jint} associated with them. This is mirrored on the level of the metric: for $\omega_x=\omega_y$, both metrics coincide, and 
\begin{equation}\label{metricscirc}
    g^{ij}_{\rm int}=g^{ij}_{\rm orb}=\frac{1}{2\ell_B^2}\delta^{ij}.
\end{equation}
In particular, this means that since $K_{\rm orb}$ is valid for the edge spin in the circular limit (c.f. Sec. \ref{sec:edgenonint}), $K_{\rm int}$ has to be as well. That this is true will be seen in Sec. \ref{sec:edgeint}.

\subsection{Edge spin using the many-body generator of rotations}\label{sec:edgeint}

\begin{figure*}[t]
\includegraphics[scale=1.2]{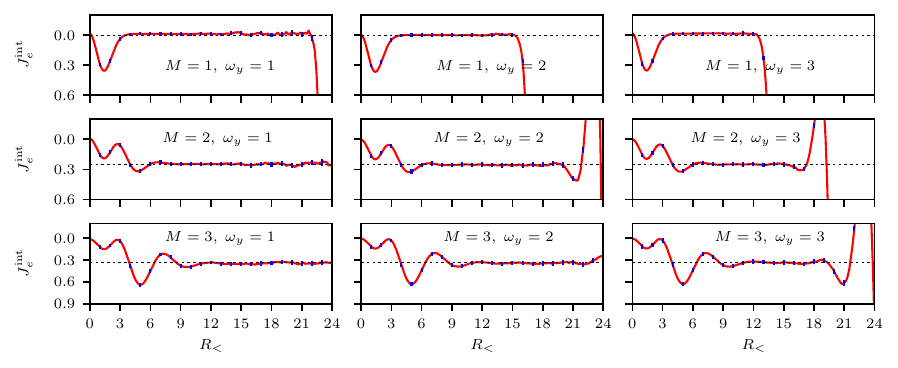}
\centering
  \caption{Edge spins $J_e^{\rm int}$ for quantum Hall droplets with $\nu=1/M$ for $M=1,2,3$, as computed using \eqref{Jeint} with $\hbar$ set to $1$. For each filling, the confinement has $\omega_x=1$ and $\omega_y=1,2$ or $3$. The number of particles is $N=300$. Throughout, the integer $N=300$ has been subtracted, and lengths are in units of $\ell_B$. Edge spins are plotted as functions of the inner cutoff $R_{<}$. The expected spin value $J_e=-\frac{1}{2}(1-\nu)$ (dashed line) is found in the plateau regions, away from the quasihole at the origin (leftmost region) and the closest parts of the edge (rightmost region). Blue error bars have been included for every $20$th plot point, i.e., at the points where $R_{<}$ takes integer values. Note that the scale differs between the third row and the other two, but stays the same within each row.}
  \label{fig:jedgeint}
\end{figure*}
Let us now compute the edge spin for the elliptic droplet using $K_{\rm int}$: 
\begin{equation}\label{Jeint}
    J_e^{\rm int}(R_{<}) = \hbar\int_0^{2 \pi} \hspace{-0.2cm} \int_{R_<}^{R_{>}}\bigg(\frac{r^2\big(1-\epsilon\cos(2\phi)\big)}{2\ell_B^2}-1\bigg)\delta\rho \;rdrd\phi.
\end{equation}
As in Sec. \ref{sec:edgenonint}, $\delta\rho=\rho_{qh}-\rho_0$ is the difference between the densities with and without a quasihole, $R_{>}$ is an outer cutoff well outside the droplet and $R_<$ is a cutoff inside the droplet, far from the edge and well outside any quasiholes in the bulk.

Edge spin values for a range of filling fractions $\nu=\frac{1}{M}=1,\frac{1}{2},\frac{1}{3}$ and confinements $\omega_x=1,\omega_y=1,2,3$ are shown in Fig.~\ref{fig:jedgeint}. From now on we set $\hbar=1$. It can be seen that all plots have their plateaux at the edge spin values (dashed lines)
\begin{equation}
J_e^{\rm int}=-\frac{1}{2}(1-\nu).
\end{equation}
These values equal $-J_{qp}$, where $J_{qp}$ is the bulk spin associated with the anyons themselves (for details on the $J_{qp}$ values and their computation, we refer again to App.~\ref{app:bulk}). They are therefore the expected edge spin values if the spin fractionalization $J_e=-J_{qp}\;\mathrm{mod}\;1$ is to hold. Given that this holds for all the different confinements and filling fractions in Fig.~\ref{fig:jedgeint}, we conclude that $K_{\rm int}$ in Eq.~\eqref{intkernel} is a suitable version of the straight-edge kernel of Ref.~\cite{fagerlund2025spin} for systems with ellipse-shaped confinement. Additionally, we conclude that spin fractionalization holds also for ellipse-shaped quantum Hall droplets. In particular, it holds in the special case of a circular edge, where $\omega_y/\omega_x=1$; see the leftmost column of Fig.~\ref{fig:jedgeint}. This suggests that spin fractionalization is not a phenomenon unique to the straight edge as studied in Ref.~\cite{fagerlund2025spin} but rather a more robust effect. 

The spin fractionalization shown numerically above can also be understood through analytical arguments.
To argue that the edge spin of the elliptical droplet fractionalizes, we consider the two eigenvalue relations \eqref{eig0} and \eqref{eig1}. As argued earlier, these relations can be represented using Eq. \eqref{jgenericmb}. Therefore, the total angular momentum (with $\hbar=1$ and $\ell_B=1$) becomes
 \begin{align}\label{jintegs}
    J_{\rm tot,i} &= \int_{r=0}^{r=R_{>}}\bigg(\frac{r^2}{2}-1-\frac{\epsilon}{2}(x^2-y^2)\bigg)\rho_i \;rdrd\phi,
\end{align}
which equals $\ell_0=\frac{M}{2}N(N-1)$ for $\rho_i=\rho_0$, and $\ell_0+N$ for $\rho_i=\rho_{qh}$. Consequently, $J_{\rm tot,qh}-J_{\rm tot, 0}=N$, the number of particles. 

The difference $J_{\rm tot,qh}-J_{\rm tot, 0}$ can also be computed by taking the difference between the integrals \eqref{jintegs}, and splitting the integration interval at some radius $R$ in the unperturbed bulk region. We take $R$ far from both the origin (where there is a quasihole for $\rho_i=\rho_{qh}$) and the edge, which gives
\begin{align}\label{splitinteg}
    J_{{\rm tot,qh}}-J_{{\rm tot},0} &= \int_{r=0}^{r=R}\bigg(\frac{r^2}{2}-1-\frac{\epsilon}{2}(x^2-y^2)\bigg)\delta\rho \;rdrd\phi\nonumber\\
    &+\int_{r=R}^{r=R_{>}}\bigg(\frac{r^2}{2}-1-\frac{\epsilon}{2}(x^2-y^2)\bigg)\delta\rho \;rdrd\phi\nonumber\\
    &=J_{qp}-\frac{\epsilon}{2}\int_{r=0}^{r=R}(x^2-y^2)\delta\rho\;dxdy+J_e^{\rm int},
\end{align}
using Eqs.~\eqref{intkernel} and \eqref{Jeint}, as well as a known expression for the quasihole bulk spin (c.f. App. \ref{app:bulk} and Refs. \cite{comparin2022spin,nardin2023ssr,fagerlund2025bosonic,fagerlund2025spin}).
We remind the reader that $\delta\rho\equiv \rho_{qh}-\rho_0$. 

As argued earlier, the LHS equals the number of particles $N$. Furthermore, the integral in the last line is over a density difference in the region at and around the quasihole, but far from the edge. This density difference is circularly symmetric, regardless of the overall shape of the droplet (see Fig. \ref{fig:cross-sections}), and therefore the integral vanishes. We conclude that
\begin{equation}
    J_e^{\rm int}=-J_{qp}+N=-J_{qp}\;\mathrm{mod}\;1,
\end{equation}
since the number of particles is (of course) an integer. Because $e^{2\pi i J}=e^{2\pi i(J+N)}$, the integer part can be discarded, and is given no importance in the rest of the paper.
When computing $J_e^{\rm int}$ numerically as in Fig. \ref{fig:jedgeint}, the integer $N$ is subtracted away, since the edge spin is only defined $\mathrm{mod}\;1$.

We remark that since the $\epsilon$-dependent parts of the integral in the topmost line of Eq. \eqref{splitinteg} vanish, the bulk spin $J_{qp}$ of the quasihole attains the expected value regardless of whether we compute it using $K_{\rm int}$ or
$K_{\rm circ}$. This holds as long as the outer integration limit $R$ is far from both the quasihole and the edge. In principle, either kernel could therefore be used for the bulk spin.

Before closing this Section, we discuss the accuracy of our numerical results. Because $J_e^{\rm int}$ in Eq.~\eqref{Jeint} depends on the difference $\rho_{qh}-\rho_0$, it can be written as the difference between two integrals of the form \eqref{jintegs}, except with the lower integration limit changed to $r=R_<$. The absolute errors of these integrals add.
This means that even a rather small \textit{relative} error could in principle make the first decimal of $J_e^{\rm int}$ unreliable. When $R_<=0$, we have seen that the integral involving $\rho_0$ would give $J_{\rm tot,0}=\frac{M}{2}N(N-1)=44,850\;M$ for $N=300$ particles. An absolute deviation of (say) $0.1$ would only correspond to a relative deviation of about $2\times 10^{-6}$ for $M=1$. Despite being such a small relative error, the absolute error would still render our edge spin values useless. Although $R_< = 0$ is not on the edge spin plateau (c.f. Fig. \ref{fig:jedgeint}), similar arguments can be made for the plateau values.
The largest absolute deviations we obtained are around $1.37\times 10^{-2}$ for $\rho_0$ and around $1.66 \times 10^{-2}$ for $\rho_{qh}$.
The deviation is often noticeably smaller, especially for $\rho_0$.  Thus, we have the necessary accuracy for our edge spin values in Fig. \ref{fig:jedgeint} to be meaningful and reliable.
 
\section{Discussion and conclusions}\label{sec:concl}
We have shown that spin fractionalization continues to hold when the droplet is deformed, regardless of whether the state is interacting or not. Therefore, spin fractionalization appears to be a rather sturdy phenomenon, which is not specific to droplets with a straight~\cite{fagerlund2025spin} or circular~\cite{nardin2023ssr} edge. Instead, it seems to be related to the topological nature of QH states. 

By considering the gauge-invariant generator of rotations for the orbitals \eqref{orbs} and the spins of quasiholes in the elliptically stretched Laughlin-like states \eqref{qhfun}, we have demonstrated that the single-particle GIGR kernel also works as an edge spin kernel for non-interacting many-body states. Once interactions are properly taken into account, the edge spin kernel~\eqref{intkernel} reflects an emergent geometry, and must be derived using the operator that deforms a circular Laughlin state into an elliptic one.
In addition, our results show that although the definition of the edge spin $J_e^{\rm int}$ heavily depends on the droplet geometry, the bulk spin $J_{qp}$ is independent of it. This is expected, as bulk properties should generally be insensitive to the external potential -- see also Appendices \ref{App:BulkProperties} and \ref{app:bulk}. This difference between edge and bulk also resonates with the results reported in Ref.~\cite{moosavi2025plasma}. 

Our results suggest a strategy for generalizing the edge spin to edges which are neither straight nor circular. Hopefully, this can allow the edge spin to be adapted to even more realistic confinement potentials and interactions in the future. In particular, a tempting extension would be to confinement geometries appropriate for interferometry setups or for manipulation of quasiparticles~\cite{willett2007confinement,nakamura2020observation,willett2009measurement,nakamura2022impact,nakamura2023fabry,bartolomei2020fractional,ruelle2023comparing,ruelle2025time}.  

Our work also connects to the geometric study of the FQHE \cite{haldane2009hall,haldane2011geometrical,read2011hall,yang2012band}. The metric \eqref{gint} for a general $M\geq 1$ fluid depends on the external potential, but is \textit{not} equal to the metric expected from the single-particle orbitals \eqref{gnonint}, except in the circular limit. This suggests that the metric is a collective property of the fluid, mirroring claims in Ref. \cite{haldane2011geometrical}.  

However, unlike in previous work \cite{haldane2011geometrical}, our metric seems unrelated to any kind of anisotropic particle-particle interactions. This can be seen in the top panel of Fig.~\ref{fig:cross-sections}, where the (electron-quasihole) correlation functions are seen to be circularly symmetric, even when changing $\omega_y/\omega_x$ (and thus also the metric). The geometry of $K_{\rm int}$ also generally differs from that of the droplet itself.

Furthermore, it is interesting to compare our results to those of Ref.~\cite{perkovic2026dipole}, where the authors study the 
edge dipole moment of FQH fluids. In particular, they consider the integer part of the edge dipole moment of both model and non-model FQH states on the cylinder, and compare their results to the predictions of~\cite{park2014dipole} (note that it was shown in Refs.~\cite{zaletel2013topfqh, zaletel2015idmrgfqh} that the fractional part of the edge dipole moment is fixed by the topological properties of the FQH fluid). 
The conclusion of Ref.~\cite{perkovic2026dipole} is that their results are only in accordance with the predictions of~\cite{park2014dipole} under special circumstances, including model states, such as the Laughlin states we consider. Thus, our results are in accordance with both~\cite{park2014dipole} and~\cite{perkovic2026dipole}. Even though we are mainly interested in the fractional part of the edge spin, it is still interesting to investigate if the angular momentum kernel we propose leads to the expected result (both the integer and fractional parts) for non-model states, such as the Jain $\nu = \frac{2}{5}$ state~\cite{jain1989cf}. Such calculations can be performed on the cylinder using MPS, or in the elliptical geometry using Monte Carlo techniques~\cite{jain1997jainkamilla}.

A promising future direction is to use the spin fractionalization shown in Ref.~\cite{fagerlund2025spin} and generalized in the present work to infer bulk spins in QH systems from edge properties. This would require a technique for sensitive measurement of the edge structure. A candidate method is the one in Ref.~\cite{unal2025circular}, based on circular dichroic response of the edge. Because our spin fractionalization is really fractionalization between a quantity dependent on the geometry of the potential ($J^{\rm int}_e$) and a quantity independent of it ($J_{qp}$), this proposal would mean using a quantity whose definition depends on the specifics of the experimental setup ($J_e^{\rm int}$) to draw conclusions about a quantity independent of most such details ($J_{qp}$). Furthermore, one could in principle tailor the confinement to make $J_e^{\rm int}$ as easy as possible to measure, in order to facilitate measurement of the anyon spin $J_{qp}.$

\section{Acknowledgments}
The authors wish to thank Christopher Ekman, Benoit Estienne, Hans Hansson, Per Moosavi, Blagoje Oblak and Steve Simon for insightful comments and interesting discussions. We thank Blagoje Oblak for useful comments on the manuscript.
A.F. also wants to thank the LPTMS at Université Paris-Saclay for the hospitality. This research utilized the Sunrise HPC facility supported by the Technical Division at the Department of Physics, Stockholm University.
This work is also part of HQI (www.hqi.fr/en/) initiative and is supported by France 2030 under the French National Research Agency grant number ANR-22-PNCQ-0002.

\bibliography{refs}

\appendix

\section{The Landau problem with an ellipse-shaped confining potential}\label{App:Landau}
In this Appendix, we provide some details omitted in Sec. \ref{sec:orbs}.

We wish to minimize the potential energy \eqref{vlllcoord}. The first term in \eqref{vlllcoord} can be expressed in terms of creation and annihilation operators $b^{\dagger},b$ within the LLL as $\mu\omega_x\omega_y\ell_B^2\big(b^{\dagger}b+\frac{1}{2}\big)$, where 
\begin{align}\label{b}
    b&=\frac{1}{\sqrt{2\omega_x\omega_y\ell_B^2}}(\omega_xR_x-i\omega_yR_y)\nonumber\\
    &=\frac{1}{\sqrt{2\omega_x\omega_y}}\bigg[(\omega_x-\omega_y)\bigg(\frac{1}{4}\frac{z}{\ell_B}-\ell_B\frac{\partial}{\partial \bar{z}}\bigg)\nonumber\\&\;\;\qquad\qquad+(\omega_x+\omega_y)\bigg(\frac{1}{4}\frac{\bar{z}}{\ell_B}+\ell_B\frac{\partial}{\partial  z}\bigg)\bigg],
\end{align}
for $\bar{z}=x-iy$. The operators are normalized so that $[b,b^{\dagger}]=1$. All in all, we have  
\begin{equation}\label{vlll}
\tilde{V}_{\rm ell}=\mu\omega_x\omega_y\ell_B^2\bigg(b^{\dagger}b+\frac{(\omega_x+\omega_y)^2}{4\omega_x\omega_y}\bigg).
\end{equation} 
In order to minimize the potential \eqref{vlll}, we require that our ground state $\psi_0$ obeys $b\psi_0=0$ in addition to $a\psi_0=0$, giving 
\begin{equation}\label{gs}
    \psi_0 = \sqrt{\frac{\sqrt{1-\epsilon^2}}{2\pi \ell_B^2}}e^{-\frac{|z|^2}{4\ell_B^2}}e^{\epsilon\frac{z^2}{4\ell_B^2}}.
\end{equation}

By applying the raising operator $b^{\dagger}$ (i.e. the adjoint of \eqref{b}), we obtain the higher eigenstates $\psi_m(z,\bar{z})=\braket{z,\bar{z}}{m}=\bra{z,\bar{z}}\frac{(b^{\dagger})^m}{\sqrt{m!}}\ket{0}$ of the potential, for which 
\begin{equation}\label{mquant}
b^{\dagger}b\psi_m=m\psi_m,
\end{equation}
implying that 
\begin{equation}\label{poteigen}
    \tilde{V}_{\rm ell}\psi_m = E_m\psi_m,
\end{equation}
for a potential energy $E_m=\mu\omega_x\omega_y\ell_B^2\big(m+\frac{(\omega_x+\omega_y)^2}{4\omega_x\omega_y}\big)$. The eigenstates $\psi_m$ are those described by \eqref{orbs} and \eqref{coeffs}. In particular, \eqref{gs} can be seen to correspond to the special case of $m=0$ in \eqref{orbs} and \eqref{coeffs}, since $H_0(z)=1$.

 We remark that the equality \eqref{mquant} holds by construction, since $b^{\dagger},b$ obey standard relations for the creation and annihilation operators of the harmonic oscillator. However, \eqref{mquant} can also be derived using  \eqref{b}, its adjoint operator, the orbitals \eqref{orbs}, and the standard relation $2(m-1)H_{m-2}(\zeta)=2\zeta H_{m-1}(\zeta)-H_m(\zeta)$ for Hermite polynomials. 

\section{Bulk properties}
\label{App:BulkProperties}

Here, we study the bulk properties of our state \eqref{lllfun} and its quasiholes, as described by \eqref{qhfun}. In so doing, we show that the properties of the state are as one would expect. For instance, it will be seen that it has the same filling fraction as the conventional, circularly symmetric Laughlin state. The quasiholes of the two states also have the same density profiles, which is evidence that \eqref{qhfun} has the same topological properties as the Laughlin state, and can be regarded as a Laughlin state reshaped by an elliptic confining potential. 

Let us examine the properties of the states \eqref{lllfun} and \eqref{qhfun} numerically using Monte Carlo integration. We begin with the density profiles, which are of particular interest due to their relation to e.g. spins and braiding properties of quasiholes \cite{comparin2022spin,fagerlund2025bosonic,fagerlund2025spin,gattu2026shape}. The density profiles were shown already in Fig. \ref{fig:shifted}, which contains quasihole density profiles in elliptic droplets with various filling fractions and confining potentials, as well as a droplet without quasiholes. It can be seen that for $\omega_y>\omega_x$, the droplet is elongated in the $x$ direction. Additionally, the quasiholes of \eqref{qhfun} are circular even for an elliptic confining potential, even if the droplet itself is not. This circular shape is preserved away from the origin, provided that the anyons are sufficiently far from the edge. This also holds when there are multiple quasiholes, and when two quasiholes coincide, as seen for the example droplet in Fig. \ref{fig:shifted}.  

There are two reasons for why the quasihole does not reflect the shape of the potential. Firstly, flux insertion in a QH sample leads to a circular electric field $E_{\phi}$, which give a purely radial current $I_r=\sigma_{r\phi}E_{\phi}$. Because the current is purely radial, the quasihole has to be rotationally symmetric. Secondly, one may view the rotational symmetry as a manifestation of screening in the bulk of the elliptically shaped FQH droplet. 

\begin{figure}
    \centering
    \includegraphics[width=\linewidth]{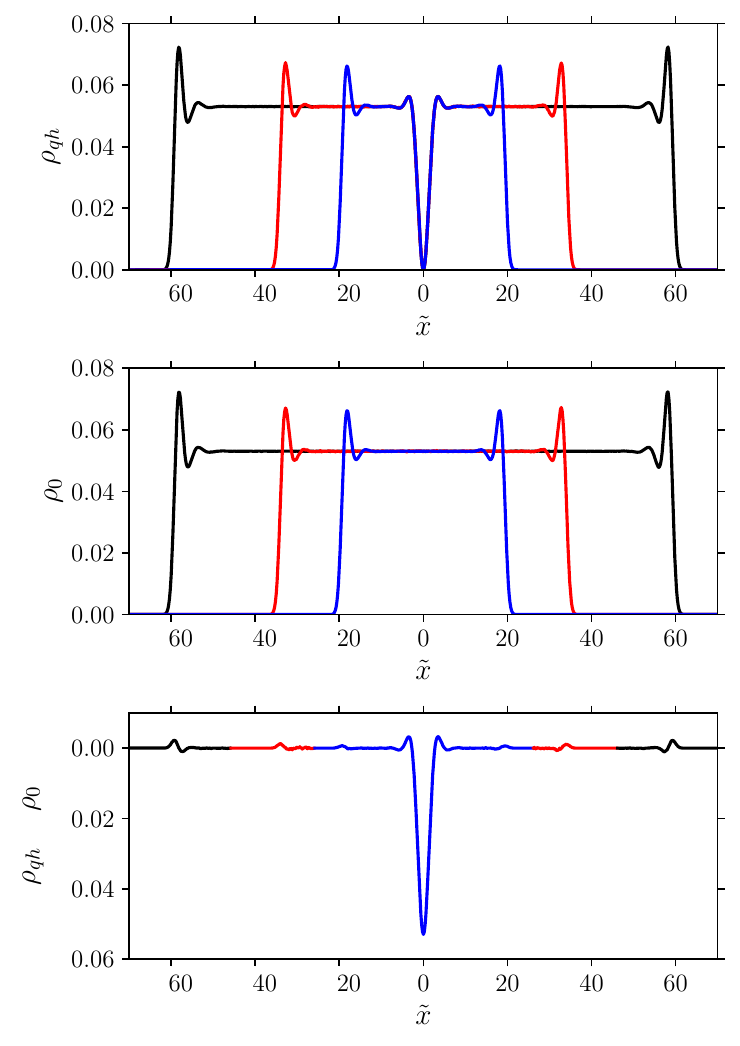}
    \caption{Upper panel: cross-sections of the density $\rho_{qh}$ with one quasihole, for $N=200$ electrons, filling fraction $\nu=1/3$, and bulk density $\frac{\nu}{2\pi}\approx 0.053$. The potential has parameters $\omega_x=\omega_y=1$ (red curve) and $\omega_x=1,\omega_y=3$ (black and blue curves). The black (blue) curve is a cross-section along the line $y=0$ ($x=0$) where the droplet is the widest (the most narrow). The coordinate $\tilde{x}=x$ for the black curve and $\tilde{x}=y$ for the others. Middle panel: cross-sections of the densities $\rho_0$ for the same systems, along the same directions. Lower panel: the density difference $\rho_{qh}-\rho_0$ between the two plots. We note that $\rho_{qh}-\rho_0$ only deviates from $0$ at the quasihole and at the edge, which is pushed out at the insertion of the quasihole. The quasihole profiles overlap for all three curves. When one curve has $\rho_{qh}-\rho_0$ visibly nonzero at the edge, the other curves have not been plotted for that region, in order to improve readability.}
    \label{fig:cross-sections}
\end{figure}

In Fig. \ref{fig:cross-sections}, we compare a quasihole profile in an elliptic droplet to the corresponding profile in a circular droplet. It is again clear that the shape of the quasihole is not affected by the potential geometry, and that even if the underlying orbitals are elliptic (see Fig. \ref{fig:orbex}), the many-body state preserves the circular shape of the quasihole.

Another observation which can be made from Fig. \ref{fig:cross-sections} is that the bulk density of our state coincides with that of the circularly symmetric Laughlin state, i.e. $\frac{\nu}{2\pi}$. This holds both when a quasihole is present and when it is not, provided that the density is measured in the unperturbed bulk sufficiently far away from the quasihole. 

Because the shape of the quasihole is unchanged, the charge and bulk spin associated with it also take the expected values. For a more detailed discussion and the appropriate definitions, the reader is referred to Appendix \ref{app:bulk}. 

Although the bulk behavior is unchanged as the droplet is stretched, the edge behavior changes. From the two upper panels of Fig. \ref{fig:cross-sections}, 
we note that the edge of the droplet experiences the same kind of angle-dependent variation of the density as the orbitals do, with a higher maximum density in the direction of the long axis; c.f. Fig. \ref{fig:orbex}. This is because the proximity to the outer edge of the droplet means that there is too little fluid further out to screen off the influence from the potential. The many-body density then ends up resembling the single-particle orbitals, which have their shape determined by the confinement.

\section{The bulk spin}\label{app:bulk}
We now show that the quasiholes of the state \eqref{qhfun} have the same spins $J_{qh}$ and charges $Q_{qh}$ as the circular Laughlin state for the same filling fraction $\nu=1/M$. These results further emphasize that our elliptic Laughlin state has the same bulk properties as the well-studied circular Laughlin state. 

\begin{figure}
    \centering
    \includegraphics[width=\linewidth]{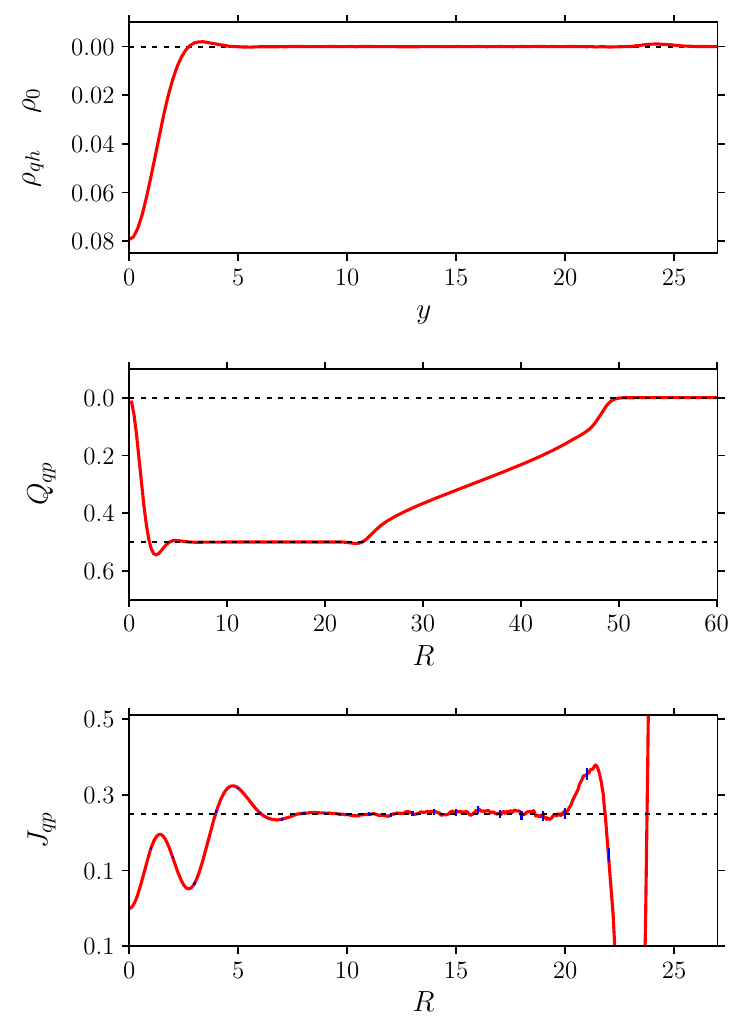}
    \caption{Upper panel: cross-section of the density difference $\rho_{qh}-\rho_0$ for a quantum Hall fluid with filling $\nu=1/2$ and potential parameters $\omega_x=1,\omega_y=2$, along the short axis of the droplet (i.e. the $y$ axis). The number of bosons is $N=300$, and distances are in units of the magnetic length $\ell_B$. Middle panel: the density difference $\rho_{qh}-\rho_0$ integrated over area as a function of outer radial cutoff $R$, see \eqref{charge}. This gives the quasihole charge $Q_{qp}=-\nu=-\frac{1}{2}$ (lower dashed line) in units where the electron charge is $+1$. When the integration reaches the most narrow edge, the value deviates from the plateau value, eventually settling at $0$ (upper dashed line) once the entire edge charge $Q_e=-Q_{qp}=\frac{1}{2}$ has been integrated over. Lower panel: the corresponding bulk spin $J_{qp}$. The expected bulk spin value $J_{qp}=\frac{1}{2}(1-\nu)=\frac{1}{4}$ (dotted line) is approached on the plateau centered around $R\approx 11\;\ell_B$. This occurs generically in the bulk region, given a sufficient distance from the quasihole (centered at $R=0$), and from the narrowest parts of the edge (around $R\approx 20\;\ell_B$), both of which have densities deviating from the bulk value. We remark that the leftmost data point for $Q_{qp}$ does not correspond to $R=0$, since the polar bins used for the Metropolis sampling in that panel cover $r$ ranges of the form $r\in [n\; dr,(n+1) dr)$ for integers $n\geq 0$, with $dr=0.25$ in units where $\ell_B=1$. Thus, the innermost bin corresponds to integration over $r\in [0,0.25)$.}
    \label{fig:jbulk}
\end{figure}

For computing the bulk spin, it has been shown that Eq.~\eqref{bulkkernel} is the suitable kernel to use \cite{comparin2022spin,nardin2023ssr}. Indeed, this kernel remains valid not just for a circular edge, as is expected from the discussion in Sec. \ref{lzours}, but for the bulk spin regardless of confining potential. This is due to the screening in the bulk, which makes any influence of the potential exponentially suppressed away from the edge. This can be seen from Fig.~\ref{fig:shifted}, where the quasiholes are sufficiently far from the edge to preserve their circular shape. We now demonstrate that the angle-independent kernel \eqref{bulkkernel} gives the expected bulk spins for our states \eqref{qhfun} as long as the quasihole is sufficiently far into the bulk, despite the droplet being ellipse-shaped due to the external potential.
Numerical results are given in Fig. \ref{fig:jbulk},
\begin{figure*}[ht]
    \centering
    \includegraphics[scale=1.2]{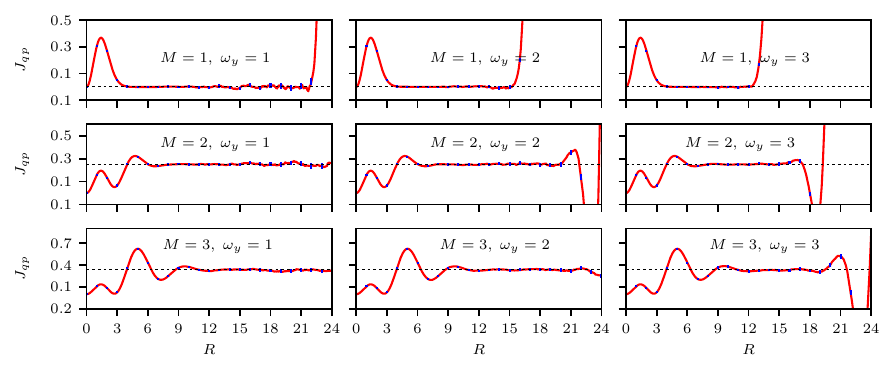}
    \caption{Bulk spins $J_{qp}(R)$ for quantum Hall droplets with $\nu=1/M$ for $M=1,2,3$, as computed using \eqref{bulkspin}. For each filling, the confinement has $\omega_x=1$ and $\omega_y=1,2$ or $3$. The number of particles is $N=300$. Lengths are in units of $\ell_B$, and we have set $\hbar=1$. Bulk spins are plotted as functions of outer cutoff radius $R$. The expected spin value $J_{qp}=\frac{1}{2}(1-\nu)$ (dashed line) is found in the plateau regions, away from the quasihole at the origin (leftmost region) and the closest parts of the edge (rightmost region). Blue error bars have been included for every $20$th plot point. Note that the scale of the vertical axis differs between different rows but stays the same within each row.}
    \label{fig:jqp}
\end{figure*}
\noindent where the quasihole bulk spin 
\begin{equation}\label{bulkspin}
    J_{qp}(R) = \hbar\int_0^{2\pi}\int_0^R \bigg(\frac{r^2}{2\ell_B^2}-1\bigg)\delta\rho(r,\phi) rdrd\phi
\end{equation}
has been computed for various values of the outer radial cutoff $R$, with $\hbar=1$ and $\ell_B=1$. Here, $R$ should be taken in the unperturbed bulk region well outside the quasihole but still far from the edge, for the integration to cover all the quasihole structure but not be affected by the edge. The density difference $\delta\rho=\rho_{qh}-\rho_0$, plotted in the uppermost panel of Fig. \ref{fig:jbulk}, is the difference between the density $\rho_{qh}$ of a system with one quasihole and the density $\rho_0$ of the same system without any quasiholes.  Coordinates are chosen such that $r=0$ is at the center of the quasihole. It can be seen by comparing the panels that the quasihole charge
\begin{equation}\label{charge}
    Q_{qp}(R)= \int_0^{2\pi}\int_0^R \delta\rho(r,\phi) rdrd\phi
\end{equation}
and the quasihole spin $J_{qp}$ both reach their expected values (marked by dashed lines) even if the droplet is elliptic in shape. This is shown in the middle and lower panels of Fig. \ref{fig:jbulk}, respectively. Both $Q_{qp}$ and $J_{qp}$ reach the same plateaux as for the circular Laughlin droplet when the integration cutoff $R$ lies in the unperturbed bulk region. In other words, these bulk observables are unaffected by the confinement. 

As outlined above, we compute the bulk spin $J_{qp}$ for a variety of states ($M=1,2,3$), each subject to a few different confinement potentials ($\omega_y/\omega_x=1,2,3$). All of these calculations are done for $N=300$ particles. The results are shown in Fig. \ref{fig:jqp}. It is seen that the bulk spin is stable against variations in both the confinement and the properties of the state -- including whether the state is interacting or not.

The spin results of Figs. \ref{fig:jbulk} and \ref{fig:jqp} are compatible with the relation postulated in \cite{fagerlund2025spin}, where inserting an integer number $p$ flux quanta into an arbitrary QH fluid gives an Abelian quasihole with spin 
\begin{equation}\label{jqpexpect}
J^{(p)}_{qp}=-\frac{1}{2\nu}\big(q_{qp}^{(p)}\big)^2-\frac{\mathcal{S}}{2}q_{qp}^{(p)}. 
\end{equation}
Here, the quasihole charge is $q_{qh}^{(p)}=-p\nu$ in units of the electron charge, and $\mathcal{S}$ is the Wen-Zee shift of the state \cite{wen1992shift}. 
For the Laughlin state, $\mathcal{S}=1/\nu$. If the above relations are valid also for our Laughlin-like state \eqref{qhfun}, a single flux quantum in this state is expected to give a bulk spin $J_{qp}^{(1)}=\frac{1}{2}(1-\nu)$ (see also Refs. \cite{comparin2022spin,trung2023ssr}). In fact, this is the expected value plotted as a dashed line in the lowermost panel of Fig. \ref{fig:jbulk}, as well as in Fig. \ref{fig:jqp}.

\section{More details on the Monte Carlo error analysis}\label{app:metropolis}
Here, we give further details on the Monte Carlo computations underlying Figs. \ref{fig:jedgenonint}, \ref{fig:jedgeint} and \ref{fig:jqp}. In these figures, the system has $N=300$ particles. The number of suggested (i.e. not necessarily accepted) Metropolis steps in each case varies somewhat, but is between $6\times 10^9$ and $1\times 10^{10}$ steps per particle. One percent of these is discarded as burn-in.

The computations have been performed by subdividing the Metropolis steps into $N_{\rm batches}$ separate batches, with each batch providing an estimate of the density with or without a quasihole. The subdivision is done to obtain effectively independent samples from the original, correlated Markov chain data \cite{krauthstatistical}, which can then be used for error estimates. The spin has been computed batchwise, with each batch of data for $\rho_{qh}$ paired together with one for $\rho_0$. For each $R_{<}$ or $R$, the spin value in Fig. \ref{fig:jedgenonint}, \ref{fig:jedgeint} or \ref{fig:jqp} is then the mean of the individual spin estimates, and the error is the (sample) standard deviation of the mean of our effectively independent samples.

\begin{figure}
     \centering
     \includegraphics[width=\linewidth]{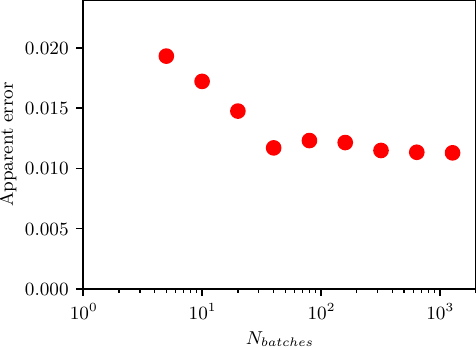}
     \caption{The effect of number of batches on the apparent error of the edge spin, for $\omega_y/\omega_x=2,M=2$. The number of Metropolis moves per particle is $N_{moves}/N=8\times 10^9$, with $N=300$. One percent is discarded as burn-in, and the data from the remaining moves is divided evenly between the different batches. We average the standard errors across the plateau points in the $M=2,\omega_y=2$ panel of Fig. \ref{fig:jedgeint}, from $R_{<}=10$ to $R_{<}=15$, in units of $\ell_B=1$. Taking the averages between batches pairwise gives half as many batches, and a new error estimate is performed. This is iterated, giving error estimates for $5\times 2^k$ batches, with $k=0,1,...,8$.}
     \label{fig:batching}
 \end{figure}

In order to find out how many batches should be used, we compute the apparent error of the plateau value of $J_{e}^{\rm int}$, for a variety of different $N_{\rm batches}$, and examine how the error depends on $N_{\rm batches}$. This is done for the case $M=2,\omega_y=2$. The result is shown in Fig.~\ref{fig:batching}. For $N_{\rm batches}$ sufficiently large, the error estimate should be reliable: it does not vary much, and in the regime shown there are still sufficiently many samples per batch for the different batches to be effectively independent. Moving left in the figure, the faster variation for $N_{\rm batches}\leq 40$ may be due to fewer samples (i.e. batches) being used to estimate the error (which is their standard deviation). Thus, we have opted to use $N_{\rm batches}=320$, to ensure that our error estimate is well away from the region where it begins to fluctuate due to there being too few batches. This is done to make the error estimate more reliable.  

 \begin{figure}
     \centering
     \includegraphics[width=\linewidth]{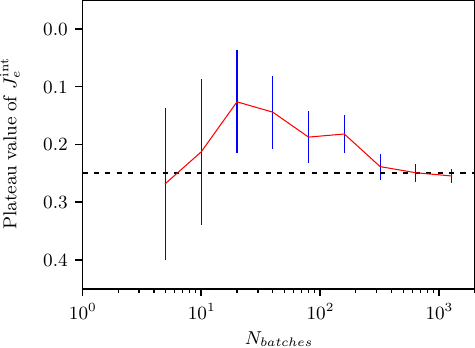}
     \caption{Mean $J_e^{\rm int}$ value across the plateau, for a droplets with $N=300$ particles, $\omega_y/\omega_x=2$, and $M=2$. The full data is $8\times 10^9$ Metropolis moves per particle, with one percent discarded as burn-in. These data points have been divided across $1280=5\times 2^8$ batches, and $5\times 2^k$ of these batches have been included for each data point, with $k=0,1,...,8$. Each data point shows the average of $J_e$ for $R_{<}$ across the plateau, from $R_{<}=10$ to $R_{<}=15$, with $\ell_B=1$. The error bar shows the standard deviation of the mean, averaged between all the plateau points in the interval. The dotted line shows the expected value $J_e^{\rm int}=-\frac{1}{4}$.}
     \label{fig:convergence}
 \end{figure}

Furthermore, we include some more detailed edge spin convergence data in Fig. \ref{fig:convergence}, still for the case $M=2,\omega_y=2$. There, $J_e^{\rm int}$ values are shown averaged across the plateau for the $M=2,\omega_y=2$ panel in Fig. \ref{fig:jedgeint}, as a function of the amount of data included in the computation. The Metropolis data for estimating the density has been divided into $1280=5\times 2^8$ batches of fixed and equal size. The horizontal axis shows how many of those batches have been included when computing $J_e^{\rm int}$, which is found as the mean value across the different batches. The error bar shows the standard deviation of that mean across the batches used. It is seen that the expected plateau value $J_e^{\rm int}=-\frac{1}{4}$ is never much more than one standard deviation of the mean away. As more data is used, the error bars shrink, and the mean stabilizes around $J_e^{\rm int}=-\frac{1}{4}$. This confirms that $J_e^{\rm int}$ indeed takes the expected value.

Finally, we mention that the random numbers for the Metropolis sampling are supplied using an xoshiro algorithm \cite{blackman2021scrambled}. Our acceptance ratios vary between runs, but are typically around $0.5.$

\section{More details on the deformation of the single-particle orbitals}\label{app:orbdefdetails}
In this Appendix, we derive the eigenstate deformation described in Sec. \ref{Sec:Def:SP:Orbitals}. 

Let us begin with a rotationally symmetric system $\omega_x=\omega_y$, and express $\Lambda_2$ in terms of the harmonic oscillator operators $b,b^{\dagger}$ in the circular limit, where
\begin{equation}\label{bbdagcirc}
    b=\sqrt{2}\bigg(\frac{1}{4}\frac{\bar{z}}{\ell_B}+\ell_B\frac{\partial}{\partial z}\bigg),\quad b^{\dagger}=\sqrt{2}\bigg(\frac{1}{4}\frac{z}{\ell_B}-\ell_B\frac{\partial}{\partial \bar{z}}\bigg).
\end{equation}
It can be shown from Eqs.~\eqref{bbdag} and \eqref{Eq:Lambda2:SP} that
\begin{equation}
    \Lambda_2=\frac{i}{2}\bigg(b^2-(b^{\dagger})^2\bigg).
\end{equation}
We also know that in the circular limit, the orbitals are the usual LLL angular momentum eigenfunctions mentioned above, which we normalize as
\begin{equation}\label{symorbs}
    \phi_m(z,\bar{z})\equiv\frac{1}{\sqrt{2\pi \ell_B^2}}\frac{1}{\sqrt{2^m m!}}\bigg(\frac{z}{\ell_B}\bigg)^me^{-\frac{|z|^2}{4\ell_B^2}}.
\end{equation}
One easily verifies that these are of the form 
\begin{equation}
    \phi_m(z,\bar{z}) = \bra{z,\bar{z}}\frac{(b^{\dagger})^m}{\sqrt{m!}}\ket{0},
\end{equation}
where $\ket{0}$ denotes the ground state of the $b,b^{\dagger}$ oscillator. This state is given in position space by
\begin{equation}
    \braket{z,\bar{z}}{0}=\frac{1}{\sqrt{2\pi \ell_B^2}}e^{-\frac{|z|^2}{4\ell_B^2}}.
\end{equation}
Thus, deforming $\phi_m$ amounts to computing 
\begin{align}\label{SbSS0}
    S\frac{(b^{\dagger})^m}{\sqrt{m!}}\ket{0} &= S\frac{(b^{\dagger})^m}{\sqrt{m!}}S^{\dagger} S\ket{0}\nonumber\\
    &= \frac{1}{\sqrt{m!}}\big(S b^{\dagger}S^{\dagger}\big)^m S\ket{0}.
\end{align}
Using Campbell's lemma \eqref{campbell} with $A=-i\lambda \Lambda_2 = \frac{\lambda}{2}\big(b^2-(b^{\dagger})^2\big)$, one finds that
\begin{align}\label{transopb}
    Sb^{\dagger} S^{\dagger}&=e^{-i\lambda\Lambda_2}b^{\dagger}e^{i\lambda\Lambda_2}\nonumber\\
    &=\sum_{k=0}^{\infty}\frac{\lambda^{2k}}{(2k)!}b^{\dagger}+\sum_{k=0}^{\infty}\frac{\lambda^{2k+1}}{(2k+1)!}b\nonumber\\
    &=\cosh(\lambda)b^{\dagger}+\sinh(\lambda)b.
\end{align}
We now compute $S\ket{0}$. A factorized expression for a squeezing operator on the same form as our operator $S$ in \eqref{squeeze} is provided in Ref. \cite{raffa2022group} (see also \cite{agarwal2012quantum}). Using this expression for real $\lambda$, we obtain
\begin{align}\label{factorize}
    S = e^{-\mathrm{sgn}(\lambda)\tanh|\lambda|\frac{(b^{\dagger})^2}{2}}e^{-\ln\big(\cosh|\lambda|\big)\big(b^{\dagger}b+\frac{1}{2}\big)}e^{\mathrm{sgn}(\lambda)\tanh|\lambda|\frac{b^2}{2}}.
\end{align}
Letting this act on $\ket{0}$, we obtain\footnote{A highly similar transformation is found in Ref. \cite{crepel2023geometric}.}
\begin{align}
    S\ket{0}&=\frac{1}{\sqrt{\cosh|\lambda|}}e^{-\mathrm{sgn}(\lambda)\tanh|\lambda|\frac{(b^{\dagger})^2}{2}}\ket{0}\nonumber\\
    &=\frac{1}{\sqrt{\cosh|\lambda|}}e^{-\mathrm{sgn}(\lambda)\tanh|\lambda|\frac{z^2}{4\ell_B^2}}\ket{0},
\end{align}
using the expression \eqref{bbdagcirc} in the second equality. We may pick $2\lambda=\ln(\omega_x/\omega_y)$. Doing so gives $-\mathrm{sgn}(\lambda)\tanh|\lambda| = \frac{\omega_y-\omega_x}{\omega_y+\omega_x} =\epsilon$ and $\cosh|\lambda|=\frac{\omega_x+\omega_y}{2\sqrt{\omega_x\omega_y}}$. This holds for both types of nontrivial deformations, $\omega_x>\omega_y$ and $\omega_y>\omega_x$, as well as for the trivial (i.e. identity) deformation with $\omega_x=\omega_y$. We obtain that
\begin{align}\label{S0}
    S\ket{0}=\frac{1}{\sqrt{2\pi \ell_B^2}}\sqrt{\frac{2\sqrt{\omega_x\omega_y}}{\omega_x+\omega_y}}e^{-\frac{|z|^2}{4\ell_B^2}}e^{\frac{\epsilon z^2}{4\ell_B^2}}.
\end{align}
Let us now put the pieces together. From Eqs.~\eqref{SbSS0}, \eqref{transopb} and \eqref{S0}, we have that
\begin{align}\label{sphi2}
    S\phi_m = \sqrt{\frac{\sqrt{1-\epsilon^2}}{2\pi \ell_B^2}}\frac{1}{\sqrt{m!}}\big(&\cosh(\lambda)b^{\dagger}\nonumber\\
    +&\sinh(\lambda)b\big)^me^{-\frac{|z|^2}{4\ell_B^2}}e^{\frac{\epsilon z^2}{4\ell_B^2}},
\end{align}
after using that $\frac{2\sqrt{\omega_x\omega_y}}{\omega_x+\omega_y}=\sqrt{1-\epsilon^2}.$ We note that, for some differentiable function $f(z)$ depending on $z$ but not on $\bar{z}$, 
\begin{equation}\label{bdagf}
    b^{\dagger} \bigg(f(z)e^{-\frac{|z|^2}{4\ell_B^2}}e^{\frac{\epsilon z^2}{4\ell_B^2}}\bigg) = \frac{z}{\sqrt{2}\ell_B}f(z)e^{-\frac{|z|^2}{4\ell_B^2}}e^{\frac{\epsilon z^2}{4\ell_B^2}},
\end{equation}
and that
\begin{equation}\label{bf}
    b \bigg(f(z)e^{-\frac{|z|^2}{4\ell_B^2}}e^{\frac{\epsilon z^2}{4\ell_B^2}}\bigg) = \bigg(\frac{\epsilon z}{\sqrt{2}\ell_B}f(z)+\sqrt{2}\ell_B\frac{\partial f}{\partial z}\bigg)e^{-\frac{|z|^2}{4\ell_B^2}}e^{\frac{\epsilon z^2}{4\ell_B^2}}.
\end{equation}
With $\cosh(\lambda)= \frac{\omega_x+\omega_y}{2\sqrt{\omega_x\omega_y}}$ and $\sinh(\lambda) = -\frac{\omega_y-\omega_x}{2\sqrt{\omega_x\omega_y}}$, as well as the definition \eqref{coeffs} of $\epsilon$, it follows that $\cosh(\lambda)+\epsilon \sinh(\lambda)=\frac{2\sqrt{\omega_x\omega_y}}{\omega_x+\omega_y}$. Thus, \eqref{bdagf} and \eqref{bf} imply that
\begin{align}
    &\big(\cosh(\lambda)b^{\dagger}
    +\sinh(\lambda)b\big)^m\bigg(f(z)e^{-\frac{|z|^2}{4\ell_B^2}}e^{\frac{\epsilon z^2}{4\ell_B^2}}\bigg)=\nonumber\\
    &\bigg[\bigg(\frac{z}{\ell_B}\frac{\sqrt{2\omega_x\omega_y}}{\omega_x+\omega_y}-\ell_B\frac{\omega_y-\omega_x}{\sqrt{2\omega_x\omega_y}}\frac{\partial}{\partial z}\bigg)^m \cdot f(z)\bigg]e^{-\frac{|z|^2}{4\ell_B^2}}e^{\frac{\epsilon z^2}{4\ell_B^2}}.
\end{align}
We now change variables to $\zeta$, as defined in \eqref{orbs}. We also specialize to the case $f=1$. This lets us conclude that \eqref{sphi2} can be written as
\begin{equation}
    \sqrt{\frac{\sqrt{1-\epsilon^2}}{2\pi \ell_B^2}}\frac{\epsilon^{\frac{m}{2}}}{\sqrt{2^m m!}}\bigg[\bigg(2\zeta-\frac{\partial}{\partial\zeta}\bigg)^m \cdot 1\bigg]e^{-\frac{|z|^2}{4\ell_B^2}}e^{\frac{\epsilon z^2}{4\ell_B^2}}.
\end{equation}
We recognize Rodrigues' formula~\cite{rodrigues1816} (see also, f.i.~\cite{arfken2013rodrigues}) for the Hermite polynomials. The above therefore becomes
\begin{align}
    S\phi_m(z,\bar{z})&=\sqrt{\frac{\sqrt{1-\epsilon^2}}{2\pi \ell_B^2}}\frac{\epsilon^{\frac{m}{2}}}{\sqrt{2^m m!}}H_m\big(\zeta\big)e^{-\frac{|z|^2}{4\ell_B^2}}e^{\frac{\epsilon z^2}{4\ell_B^2}}\nonumber\\
    &=\psi_m,
\end{align}
using \eqref{orbs} and \eqref{coeffs}. Thus, the circular LLL orbitals \eqref{symorbs} are mapped to the elliptic orbitals \eqref{orbs} by our operator $S$, as was to be shown.

\section{An explanation for why the single-particle generator of rotations sometimes does not apply to the many-body state}\label{app:motivation}
We now proceed to explain why the generator of rotations used for the single-particle orbitals can be used in the circular limit, and for non-circular, non-interacting ($M=1$) states, but not for non-circular $M>1$ systems. In the circular limit, it does not matter for the generator of rotations whether there are interactions ($M>1$) or not ($M=1$). In both of these cases, the state is an eigenstate of the (circular limit of the) GIGR. The reason is that after expanding the polynomial part of the wavefunction, all terms have the same total degree, as can be seen from the elementary binomial expansion 
\begin{equation}
    (z_i-z_j)^M=\sum_{k=0}^M \binom{M}{k}(-1)^k z_i^{M-k}z_j^k,
\end{equation}
where it is manifest that the total degree of each term is $M$. Therefore, the Laughlin state is an angular momentum eigenstate also for $M>1$, just like the single-particle orbitals. 

For a non-circular state, where $\epsilon\neq 0$, the above reasoning is modified. Let us consider the appropriate generalization of angular momentum to a non-interacting state without rotational symmetry, which is the total eigenvalue of the orbital index $\sum_i b^{\dagger}_ib_i$, where $b_i$ is the operator \eqref{b} acting on the coordinates $z_i,\bar{z}_i$. For the non-interacting state $M=1$, the polynomial part consists of factors
\begin{equation}
    (z_i-z_j)=\frac{1}{2}\big(H_1(z_i)-H_1(z_j)\big),
\end{equation}
where all terms (once the Gaussians are included) have the same eigenvalue of $\sum_i b_i^{\dagger}b_i$ (c.f. Eqs.~\eqref{orbs} and \eqref{mquant}). However, expanding the polynomial for the interacting state (i.e. that where $M>1$) in the basis of Hermite polynomials, we obtain factors like
\begin{align}
    (z_i-z_j)^2&=\frac{1}{2}\big(H_1(z_i)-H_1(z_j)\big)^2\nonumber\\
    &=\frac{1}{4}\bigg(H_2(z_i)+2H_0(z_i)\bigg)H_0(z_j)\nonumber\\&+\frac{1}{2}H_1(z_i)H_1(z_j)&\nonumber\\&+\frac{1}{4}H_0(z_i)\bigg(H_2(z_j)+2H_0(z_j)\bigg).
\end{align}
It is clear that the above (with the Gaussians reinstated) is not an eigenstate of $\sum_i b^{\dagger}_ib_i$. On an algebraic level, the state ceases to be an eigenstate of the single-particle generator of rotations because $H_n^M(z)\neq H_1^{Mn}(z)$. In the circular limit it continues to be an eigenstate, because the polynomial factors become $z^n$, which (trivially) obey $(z^n)^M=z^{Mn}.$

\section{More general confinement shapes}\label{sec:lzprx}
In this Appendix, we briefly investigate a bigger class of quantum Hall droplets where anisotropy is introduced by distorting the confining potential as 
    \begin{equation}\label{distort}
    V\big(r^2\big)\rightarrow V\bigg(\frac{r^2}{2f'(\phi)}\bigg) .
    \end{equation}
This class of potentials was studied in~\cite{oblak2024anisotropic}, where it was found that the probability densities associated to the orbitals $\chi_m$ are, in the large-$m$ limit, 
\begin{equation}\label{asymporbs}
    |\chi_m(r,\phi)|^2
    =\frac{1}{2\pi \ell_B^2\sigma(\phi)}\frac{e^{-2\frac{a^2}{\sigma^2}}}{(2\pi m)^{1/2}}.
\end{equation}
The parameters appearing in the expression are
\begin{equation}
   a(r,\phi)=\frac{r-\ell_B\sqrt{2mf'(\phi)}}{\ell_B\sqrt{2f'(\phi)}} 
\end{equation}
and
\begin{equation}\label{sigma}
    \sigma(\phi)=\frac{1}{\sqrt{f'(\phi)}}\sqrt{1+\bigg(\frac{f''(\phi)}{2f'(\phi)}\bigg)^2}.
\end{equation}
This result follows from computing the dominant term in a $1/m$ expansion of the wavefunction and is thus particularly useful when $m \gg 1$.

A natural definition of the integral kernel $K(x,y)$ for these orbitals is 
\begin{equation}\label{kernel}
K_{m \gg 1}(r,\phi)=\hbar \left( \frac{r^2}{2f\;'(\phi)\ell_B^2}- \frac{3}{4}\sigma^2(\phi) \right).
\end{equation}
One can verify that with this definition, and using the probability density in Eq.~\eqref{asymporbs}, the integral in Eq.~\eqref{jgeneric} evaluates to  $\hbar m$, which is here interpreted as the eigenvalue of the generalized GIGR.

Interestingly, the kernel~\eqref{kernel} shares several features with the kernel in Eq.~\eqref{ourkernel}. The most notable are that it is composed of two terms (one quadratic and one constant), where the former exactly follows the equipotential lines of the confinement.
When $f'$ in \eqref{distort} is constant, and thus the confining potential is rotationally symmetric, the equipotentials are circular and it is easy to verify that $\sigma(\phi)=1$ becomes independent of the angle $\phi$. In this circular limit, the kernel~\eqref{kernel} reduces to $\frac{r^2}{2\ell_B^2}-\frac{3}{4}$ for a rotationally invariant droplet. The additive constant thus differs from that of the exact kernel in Eq.~\eqref{bulkkernel}.
This is caused by the large-$m$ approximation made in Ref.~\cite{oblak2024anisotropic}.
This motivates us to focus our attention exclusively on the first term of the kernel $K_{m \gg 1}$.

A special case of the above distortions are the so-called “flower deformations"
\begin{equation}\label{flower}
f'(\phi)=\frac{1}{\cosh(2\lambda)+\sinh(2\lambda)\cos(k\phi)},
\end{equation}
for integer $k$~\cite{oblak2024anisotropic}. The ellipse corresponds to $k=2$, and the equipotential curves are shaped like flowers with $k$ petals for $k\geq 3$. For these deformations, the kernel \eqref{kernel} takes the form 
\begin{equation}\label{flowerkernel}
   K_{m \gg 1} =
   \frac{r^2\big[\omega_x^2+\omega_y^2+\big(\omega_x^2-\omega_y^2\big)\cos(k\phi)\big]}{4\omega_x\omega_y \ell_B^2}-\frac{3}{4}\sigma^2(\phi),
\end{equation}
provided that we set
$   \lambda=\frac{1}{2}\ln(\omega_x/\omega_y)$.

For $k=2$, the kernel $K_{m \gg 1}(x,y)$ in turn reduces to 
\begin{equation}\label{flowerkernelelliptic}
K_{m \gg 1}=\frac{\omega_x^2x^2+\omega_y^2y^2}{2\omega_x\omega_y \ell_B^2}-\frac{3}{4}\sigma^2(\phi),
\end{equation}
since $
\cos(2\phi)=\cos^2\phi-\sin^2\phi=\frac{1}{r^2}(x^2-y^2).$
We recognize the quadratic part of this kernel: it is the same as for $K_{\rm orb}(x,y)$ in \eqref{ourkernel}. 
The factor $\sigma^2(\phi)$ in Eq.~\eqref{flowerkernelelliptic} instead reads, for $k=2$, 
\begin{equation}
    \sigma^2(\phi)=\frac{\omega_x^2+\omega_y^2}{\omega_x\omega_y}-\frac{2\omega_x\omega_y}{\omega_x^2+\omega_y^2+(\omega_x-\omega_y)(\omega_x+\omega_y)\cos(2\phi)},
\end{equation}
which is manifestly independent of $\phi$ when $\omega_x=\omega_y$, but is otherwise a nontrivial $\pi$-periodic function of $\phi$.

Although deformations beyond the elliptic one will not be further studied in this work, the observations contained in this Appendix suggest a way of generalizing our discussion to a wider class of confining potentials.

\end{document}